\documentclass[pra,reprint,superscriptaddress]{revtex4-2}
\usepackage[T1]{fontenc}
\usepackage[latin9]{inputenc}
\usepackage{amsbsy}
\usepackage{amssymb}
\usepackage{esint}
\usepackage{babel}
\usepackage{graphicx}
\begin{document}
\title{Variational approach to light-matter interaction: Bridging quantum and semiclassical limits}
\author{Yiying Yan}\email{yiyingyan@zust.edu.cn}
\affiliation{Department of Physics, School of Science, Zhejiang University of Science and Technology, Hangzhou 310023, China}
\author{Zhiguo L\"{u}}\email{zglv@sjtu.edu.cn}
\affiliation{Key Laboratory of Artificial Structures and Quantum Control
(Ministry of Education), School of Physics and Astronomy,
Shanghai Jiao Tong University, Shanghai 200240, China}
\author{JunYan Luo}
\affiliation{Department of Physics, School of Science, Zhejiang University of Science and Technology, Hangzhou 310023, China}

\date{\today}
\begin{abstract}
    We present a time-dependent variational approach with the multiple Davydov $D_2$ trial state to simulate the dynamics of light-matter systems
    when the field is in a coherent state with an arbitrary finite mean photon number.
    The variational approach captures not only the system dynamics but also the field dynamics
    and is applicable to a variety of quantum models of light-matter interaction such as the Jaynes-Cummings model, Rabi model, and Dicke model,
    and is feasible to tackle the multimode quantized fields. By comparison of the variational and semiclassical dynamics of both the system and field,
    we illustrate that the variational dynamics from the quantum models agrees
    with those from the corresponding semiclassical models as long as the mean number of photons
    is sufficiently large. 
    Moreover, we illustrate that in the crossover between the quantum and semiclassical limits, the quantum corrections lead to the collapse of the oscillations in dynamics, which is absent in the semiclassical models.
    The variational approach provides a unified treatment of light-matter interaction from the quantum to the semiclassical limit.
\end{abstract}
\maketitle

\section{Introduction}

Light-matter interaction between quantum systems
and coherent fields plays a fundamental role in quantum optics~\cite{scully,cohen}, which enables
the control of the quantum system by the field or vice versa and
is highly relevant to the realization of quantum technologies
such as quantum information processing~\cite{Rivera_2020}, quantum sensing~\cite{Degen_2017,Karnieli_2023}, quantum metrology~\cite{Hoo_2011,Bai_2023}, 
quantum battery~\cite{Faezeh_2019,Zhangyy_2019,Crescente_2020,Arjmandi}
etc. Theoretically,
there exist two kinds of descriptions of the light-matter interaction.
One is based on the full quantum description with a quantized field.
The other is based on the semiclassical description where the field
is treated classically. Two paradigmatic models for such theories
are quantum and semiclassical Rabi models~\cite{Braak_2011,Braak_2016,Xie_2017}. Of particular interest,
the study of these basic models has recieved renewed attention in artificial atoms. Both
theoretical and experimental studies on the quantum Rabi model have been
extended to the so-called ultrastrong coupling regime~\cite{Yoshihara_2016,Simone_2017,Lv_2018,Frisk_Kockum_2019,Forn_D_az_2019,Le_Boit_2020}, where the coupling
constant becomes a considerable fraction of the field frequency. For
the semiclassical Rabi model, a strong driving with the Rabi frequency
being comparable to the transition frequency of the qubit can be achieved~\cite{Yoshihara_2014,Deng_2015}.
Such a strong coupling (driving) regime has potential applications
in the realization of ultrafast quantum gates~\cite{Romero_2012,Song_2016}.

Provided the field is initially in a coherent state, the quantum and
semiclassical descriptions of light-matter interaction become consistent
with each other in the semiclassical limit, that is, the mean photon
number of the field tends to infinity and the coupling constant tends
to zero while their product remains constant~\cite{scully,cohen,irish_2022}. However, in realistic
situations, the quantum system can only interact with a finite number
of photons other than an infinite number of photons. A question naturally
raises that under which condition a finite number of photons cause
a negligible deviation between the quantum and semiclassical models.
Moreover, how quantum corrections in the presence of a finite number of
photons contribute to the dynamics remains, to the best
of our knowledge, barely explored, which may be helpful
in understanding particularly the crossover between the quantum and
semiclassical limit.

Since fields are introduced in different ways in the quantum and semiclassical
Hamiltonians, the light-matter interactions in the
quantum and semiclassical limit are typically treated by different
theoretical approaches. For instance, there are a variety of theoretical
methods developed for the quantum Rabi model and its variants, e.g., von Vleck perturbation~\cite{hausinger_2010},
unitary transformations~\cite{Gan_2010,Zhang_2016,Lei_2019}, variational approach~\cite{Werther_2017,Werther_2018,Lizm_2021}, extended coherent states~\cite{chen_2012},
etc. However, the Floquet theory mainly applies to the semiclassical
models~\cite{Shirley_1965,Chu_2004}. In contrast to the full quantum models, the field dynamics
is ignored in the semiclassical models. Very recently, a so-called
photon-resolved Floquet theory has been proposed to study the field
dynamics of semiclassical models~\cite{Georg_2024}. However, such an approach is
established in the semiclassical limit and thus is inapplicable
in the crossover between the quantum and semiclassical limit.

In this work, we present a unified numerical method that combines
two sequential unitary transformations with a time-dependent variational
approach equipped with the multiple Davydov ansatz~\cite{Zhao_2022,Zhao_2023} to simulate the
dynamics of light-matter systems from quantum to semiclassical limit
when the field is initially in a coherent state. One advantage of
the variational approach is that the multiple Davydov ansatz uses
the coherent states as the bases, which effectively prevents the exponential
increase in the size of the Hilbert space due to the increase in the
number of modes, and thus is applicable to a multimode case. More
importantly, the variational approach simultaneously captures the
dynamics of both the quantum system and fields. Particularly, we illustrate
that the statistical characteristic function of the field can be
computed, which allows us to calculate the mean value and variance of
the photon numbers of the field as well as the photon-number distribution.
We also show how the field dynamics can be calculated from semiclassical
models. We apply the present approaches to
a variety of light-matter systems including the Jaynes-Cummings (JC) model, the Rabi model, and the Dicke model.
By comparing the variational dynamics from the full quantum models with that from the semiclassical models, we examine the consistency
between the quantum and semiclassical models in the presence of a large number of photons and illustrate the role of quantum corrections in the quantum-semiclassical crossover.

The rest of the manuscript is organized as follows. In Sec.~\ref{sec:mm},
we present the variational approach to treat the light-matter interaction in the presence of large mean photon numbers.
We also show how to calculate the field dynamics in the semiclassical model.
In Sec.~\ref{sec:app}, we apply the variational approach and semiclassical approach to several light-matter systems and calculate the system and field dynamics.
In Sec.~\ref{sec:con}, we draw the conclusions.

\section{Model and methods}\label{sec:mm}

The full quantum description of the interaction between a quantum system
and a multimode coherent field can be described by the following Hamiltonian
(we set $\hbar=1$ throughout this work)
\begin{equation}
H=H_{{\rm S}}+\sum_{k=1}^{N}\omega_{k}b_{k}^{\dagger}b_{k}+\sum_{k=1}^{N}\frac{g_{k}}{2}(b_{k}^{\dagger}V_{k}+b_{k}V_{k}^{\dagger}),\label{eq:Ham}
\end{equation}
where $H_{{\rm S}}$ is the free Hamiltonian of the quantum 
system and $V_{k}$ is the interaction operator acting on the Hilbert
space of the quantum system. $b_{k}$ ($b_{k}^{\dagger}$) is the
annihilation (creation) operator of the bosonic field with frequency
$\omega_{k}$. $g_{k}$ is the coupling constant. $N$ is the total
number of modes. Note that either $V_{k}\neq V_{k}^{\dagger}$ or $V_k=V_k^\dagger$ is possible,
corresponding to a rotating-wave approximation (RWA) and non-RWA Hamiltonian, respectively.

The time evolution of the composite system is governed by the time-dependent
Schr\"{o}dinger equation,
\begin{equation}
i\frac{d}{dt}|\Psi(t)\rangle=H|\Psi(t)\rangle,
\end{equation}
where $|\Psi(t)\rangle$ is the state of the total system. In this
work, we consider a factorized initial state of the total system
\begin{equation}
|\Psi(0)\rangle=|\psi(0)\rangle\otimes|\vec{\alpha}\rangle,\label{eq:inis}
\end{equation}
where $|\psi(0)\rangle$ is an initial state of the quantum system,
and $|\vec{\alpha}\rangle$ is a multimode coherent state and reads
\begin{eqnarray}
|\vec{\alpha}\rangle & = & |\alpha_{1}\rangle\otimes|\alpha_{2}\rangle\otimes\cdots\otimes|\alpha_{N}\rangle\nonumber \\
 & \equiv & \exp\left(\sum_{k=1}^{N}\alpha_{k}b_{k}^{\dagger}-{\rm h.c.}\right)|\boldsymbol{0}\rangle\equiv D(\vec{\alpha})|\boldsymbol{0}\rangle,
\end{eqnarray}
where $|\boldsymbol{0}\rangle$ is a multimode vacuum state, $D(\vec{\alpha})$
is a displacement operator, and $\alpha_{k}\equiv|\alpha_{k}|e^{-i\phi_{k}}$
is a complex number with the modulus $|\alpha_{k}|$ and phase $\phi_{k}$.

\subsection{Time-dependent variational approach}

To achieve a manageable numerical simulation, we convert the time-evolution
problem with the initial coherent state of a large number of photons
into a new time-evolution problem with an initial vacuum state. This
can be achieved by two sequential unitary transformations. First,
we transform the time-dependent Schr\"{o}dinger equation into the
interaction picture governed by the free Hamiltonian of the field
$H_{{\rm F}}=\sum_{k=1}^{N}\omega_{k}b_{k}^{\dagger}b_{k}$. In doing
so, we have a new time-evolution problem:
\begin{equation}
i\frac{d}{dt}|\tilde{\Psi}(t)\rangle=\tilde{H}(t)|\tilde{\Psi}(t)\rangle,
\end{equation}
where the Hamiltonian becomes time dependent
\begin{equation}
\tilde{H}(t)=H_{{\rm S}}+\sum_{k=1}^{N}\frac{g_{k}}{2}(V_{k}^{\dagger}b_{k}e^{-i\omega_{k}t}+V_{k}b_{k}^{\dagger}e^{i\omega_{k}t}),
\end{equation}
and the transformed wave function is related to the original one via
\begin{equation}
|\tilde{\Psi}(t)\rangle=\exp\left(iH_{{\rm F}}t\right)|\Psi(t)\rangle.
\end{equation}
The initial state in the transformed frame remains the same as that
in the original one $|\tilde{\Psi}(0)\rangle=|\Psi(0)\rangle.$ To
proceed, we apply a displacement transformation on the equation of
motion and the initial state, yielding
\begin{equation}
i\frac{d}{dt}|\Psi^{\prime}(t)\rangle=H^{\prime}(t)|\Psi^{\prime}(t)\rangle,\label{eq:eom}
\end{equation}
where the displaced wave function is related to the original one via
\begin{equation}
|\Psi^{\prime}(t)\rangle=D^{\dagger}(\vec{\alpha})\exp\left(iH_{{\rm F}}t\right)|\Psi(t)\rangle,
\end{equation}
and the transformed Hamiltonian is given by
\begin{eqnarray}
H^{\prime}(t) & = & D^{\dagger}(\vec{\alpha})\tilde{H}(t)D(\vec{\alpha})\nonumber \\
 & = & H_{{\rm S}}(t)+\sum_{k=1}^{N}\frac{g_{k}}{2}(V_{k}^{\dagger}b_{k}e^{-i\omega_{k}t}+V_{k}b_{k}^{\dagger}e^{i\omega_{k}t}),\label{eq:Hamp}
\end{eqnarray}
where
\begin{equation}
    H_{{\rm S}}(t)=H_{{\rm S}}+\sum_{k=1}^{N}\frac{\Omega_{k}}{2}[V_{k}e^{i(\omega_{k}t+\phi_{k})}+V_{k}^{\dagger}e^{-i(\omega_{k}t+\phi_{k})}],
    \end{equation}
\begin{equation}
\Omega_{k}=|\alpha_{k}|g_{k}.
\end{equation}
In doing so, we have the following initial state
\begin{equation}
|\Psi^{\prime}(0)\rangle=D^{\dagger}(\vec{\alpha})|\Psi(0)\rangle=|\psi(0)\rangle|\boldsymbol{0}\rangle,
\end{equation}
where the field is initially in the vacuum state in the transformed
frame. Note that in the semiclassical limit of $g_{k}\rightarrow0$
and $\alpha_{k}\rightarrow\infty$ but $|\alpha_{k}|g_{k}=\Omega_{k}$,
the full quantum Hamiltonian of light-matter interaction becomes
the semiclassical Hamiltonian, i.e., $H^\prime(t)\rightarrow H_{\rm S}(t)$~\cite{irish_2022}.

We now use a time-dependent variational approach to compute the dynamics
described by Eq.~(\ref{eq:eom}). The variational approach is based
on the Dirac-Frenkel time-dependent variational principle and the
multiple Davydov $D_{2}$ ansatz. The former allows us to calculate
the optimal solution to the time-dependent Schr\"{o}dinger equation
with a parameterized trial state. In this work, we use the multiple
Davydov $D_{2}$ ansatz $|D_{2}^{M}(t)\rangle$, which is suitable
for the spin-boson-like problems and is parameterized as
follows~\cite{WangLu_2017,Werther_2020,Zhao_2022,Zhao_2023,YanPRA_2023}:
\begin{equation}
|D_{2}^{M}(t)\rangle=\sum_{n=1}^{M}\sum_{j=1}^{N_{S}}A_{nj}|j\rangle|f_{n}\rangle,
\end{equation}
where $M$ is the number of the coherent states, $A_{nj}$ are time-dependent variational parameters, $\{|j\rangle|j=1,2,\ldots,N_{S}\}$
represents a set of orthornormal bases for the quantum system, and
\begin{eqnarray}
|f_{n}\rangle & = & \exp\left(\sum_{k=1}^{N}f_{nk}b_{k}^{\dagger}-{\rm h.c.}\right)|\boldsymbol{0}\rangle
\end{eqnarray}
are multimode coherent states with $f_{nk}$ being time-dependent
variational parameters. The equations of motion for variational parameters
are determined by the Dirac-Frenkel time-dependent variational principle~\cite{frenkel},
that is,
\begin{equation}
\langle\delta D_{2}^{M}(t)|[i\partial_{t}-H^{\prime}(t)]|D_{2}^{M}(t)\rangle=0,\label{eq:tdvp}
\end{equation}
where $\langle\delta D_{2}^{M}(t)|$ represents the variation of the adjoint
state. From Eq.~(\ref{eq:tdvp}), the equations of
motion for the variational parameters are simply given by
\begin{equation}
i\langle j|\langle f_{l}|\dot{D}_{2}^{M}(t)\rangle=\langle j|\langle f_{l}|H^{\prime}(t)|D_{2}^{M}(t)\rangle,\label{eq:eom1}
\end{equation}
\begin{equation}
i\sum_{j=1}^{N_{S}}A_{lj}^{\ast}\langle j|\langle f_{l}|b_{k}|\dot{D}_{2}^{M}(t)\rangle=\sum_{j=1}^{N_{S}}A_{lj}^{\ast}\langle j|\langle f_{l}|b_{k}H^{\prime}(t)|D_{2}^{M}(t)\rangle.\label{eq:eom2}
\end{equation}
These equations of motion are  a set of implicit first-order
nonlinear differential equations and can be solved by the fourth-order
Runge-Kutta algorithm~\cite{Werther_2020,YanPRA_2023}. The detailed derivation
and implementing numerical simulation of Eqs.~(\ref{eq:eom1}) and~(\ref{eq:eom2}) are presented in Appendix.

On solving the equations of motion, we can compute the quantities
of interest for both the quantum system and the field. The observable
of quantum system can be directly computed since the applied unitary transformations
do not influence the operators acting on the Hilbert space
of the matter system, e.g., the population of the state $|j\rangle$ can be given by
\begin{eqnarray}
P_{j}(t) & = & \langle D_{2}^{M}(t)|j\rangle\langle j|D_{2}^{M}(t)\rangle\nonumber \\
 & = & \sum_{l,n=1}^{M}A_{lj}^{\ast}{\cal S}_{ln}A_{nj},
\end{eqnarray}
where
\begin{equation}\label{eq:Sln}
{\cal S}_{ln}=\exp\left[\sum_{k=1}^{N}\left(f_{lk}^{\ast}f_{nk}-\frac{|f_{lk}|^{2}}{2}-\frac{|f_{nk}|^{2}}{2}\right)\right].
\end{equation}
For the field, we calculate the moment-generating function,
\begin{eqnarray}
G_{\vec{\chi}}(t) & = & \langle e^{i\sum_{k=1}^{N}\chi_{k}b_{k}^{\dagger}b_{k}}\rangle_{t}\nonumber \\
 & = & \langle D_{2}^{M}(t)|D^{\dagger}(\vec{\alpha})e^{i\sum_{k=1}^{N}\chi_{k}b_{k}^{\dagger}b_{k}}D(\vec{\alpha})|D_{2}^{M}(t)\rangle\nonumber \\
 & = & \sum_{n,l=1}^{M}\sum_{j=1}^{N_{S}}A_{lj}^{\ast}{\cal S}_{ln}A_{nj}\nonumber \\
 &  & \times\exp\left[\sum_{k=1}^{N}(e^{i\chi_{k}}-1)(\alpha_{k}^{\ast}+f_{lk}^{\ast})(\alpha_{k}+f_{nk})\right],
\end{eqnarray}
where $\chi_{k}\in[0,2\pi)$, $\langle\cdot\rangle_{t}$ represents
the average taken over the state in the laboratory frame,
and we have used the identity of the displacement operators:
\begin{equation}
D(\alpha)D(\beta)=\exp\left(\frac{\alpha\beta^{\ast}-\alpha^{\ast}\beta}{2}\right)D(\alpha+\beta).
\end{equation}
The moment-generating function contains the information of interest
for the field. Particularly, the mean photon number in the mode $k$
can be computed as
\begin{eqnarray}
n_{k}(t) & = & \langle b_{k}^{\dagger}b_{k}\rangle_{t}\nonumber \\
 & = & \left.\frac{d}{id\chi_{k}}G_{\vec{\chi}}(t)\right|_{\chi_{k}=0}\nonumber \\
 & = & \sum_{l,n=1}^{M}\sum_{j=1}^{N_{S}}A_{lj}^{\ast}{\cal S}_{ln}A_{nj}(\alpha_{k}^{\ast}+f_{lk}^{\ast})(\alpha_{k}+f_{nk}).
\end{eqnarray}
The variance of the photon number of the mode $k$ is given by
\begin{equation}
\sigma_{k}^{2}(t)=\left.\frac{d^{2}}{i^{2}d\chi_{k}^{2}}G_{\vec{\chi}}(t)\right|_{\chi_{k}=0}-n_{k}^{2}(t).
\end{equation}
Additionally, the probability distribution of the photon numbers at
time $t$ can also be computed,
\begin{equation}
p(\vec{n},t)=\frac{1}{(2\pi)^{N}}\int_{0}^{2\pi}G_{\vec{\chi}}(t)e^{-i\vec{\chi}\cdot\vec{n}}d\vec{\chi},
\end{equation}
where $\vec{n}=(n_{1},n_{2},\ldots,n_{N})$ with $n_{k}$ being a
non-negative integer that specifies the photon number of the mode
$k$.

\subsection{Dynamics in the semiclassical limit}

In this section, we compute the dynamics of the quantum system and
field in the semiclassical limit. We first compute the population
dynamics of the quantum system with Eq.~(\ref{eq:eom}). To this end,
we express the time-evolution operator $U^{\prime}(t)={\cal T}\exp\left[-i\int_{0}^{t}H^{\prime}(\tau)d\tau\right]$
with ${\cal T}$ the time-ordering operator by using the product form,
\begin{equation}
U^{\prime}(t)=U_{{\rm S}}(t)U_{{\rm I}}^{\prime}(t),\label{eq:Upt}
\end{equation}
where
\begin{equation}
U_{{\rm S}}(t)={\cal T}\exp\left[-i\int_{0}^{t}H_{{\rm S}}(\tau)d\tau\right]
\end{equation}
is the time-evolution operator for the semiclassical Hamiltonian.
$U_{{\rm I}}^{\prime}(t)$ can be determined by iteration from the
time-dependent Schr\"{o}dinger equation and is readily given
in terms of the Dyson's series, which reads
\begin{eqnarray}
U_{{\rm I}}^{\prime}(t) & = & 1-i\int_{0}^{t}H_{{\rm I}}(\tau_{1})d\tau_{1}\nonumber \\
 &  & -\int_{0}^{t}\int_{0}^{\tau_{1}}d\tau_{1}d\tau_{2}H_{{\rm I}}(\tau_{1})H_{{\rm I}}(\tau_{2})+\ldots,\label{eq:UIT}
\end{eqnarray}
where
\begin{equation}
H_{{\rm I}}(t)=\sum_{k=1}^{N}\frac{g_{k}}{2}[V_{k}^{\dagger}(t)b_{k}e^{-i\omega_{k}t}+V_{k}(t)b_{k}^{\dagger}e^{i\omega_{k}t}],
\end{equation}
\begin{equation}
V_{k}(t)=U_{{\rm S}}^{\dagger}(t)V_{k}U_{{\rm S}}(t).
\end{equation}
 With the above results at hand, we can calculate the population of
the state $|j\rangle$ as follows:
\begin{eqnarray}
P_{j}(t) & = & \langle\Psi^{\prime}(t)|j\rangle\langle j|\Psi^{\prime}(t)\rangle\nonumber \\
 & = & \langle\Psi^{\prime}(0)|U^{\prime\dagger}(t)|j\rangle\langle j|U^{\prime}(t)|\Psi^{\prime}(0)\rangle\nonumber \\
 & = & P_{j}^{{\rm sc}}(t)+P_{j}^{{\rm qc}}(t),
\end{eqnarray}
with
\begin{equation}
P_{j}^{{\rm sc}}(t)=\langle\Pi_{j}(t)\rangle_{0}, \label{eq:pjsc}
\end{equation}
\begin{widetext}

\begin{eqnarray}
P_{j}^{{\rm qc}}(t) & = & \sum_{k=1}^{N}\frac{g_{k}^{2}}{4}\int_{0}^{t}\int_{0}^{t}d\tau_{1}d\tau_{2}\langle V_{k}^{\dagger}(\tau_{1})\Pi_{j}(t)V_{k}(\tau_{2})\rangle_{0}e^{-i\omega_{k}(\tau_{1}-\tau_{2})}\nonumber \\
 &  & -{\rm Re}\sum_{k=1}^{N}\frac{g_{k}^{2}}{2}\int_{0}^{t}\int_{0}^{\tau_{1}}d\tau_{1}d\tau_{2}\langle\Pi_{j}(t)V_{k}^{\dagger}(\tau_{1})V_{k}(\tau_{2})\rangle_{0}e^{-i\omega_{k}(\tau_{1}-\tau_{2})}+O(g_{k}^{4}), \label{eq:qc}
\end{eqnarray}
\end{widetext}where $\langle\cdot\rangle_{0}$ represents the average
taken over the initial state of the quantum system $|\psi(0)\rangle$
and
\begin{equation}
\Pi_{j}(t)=U_{{\rm S}}^{\dagger}(t)|j\rangle\langle j|U_{{\rm S}}(t).
\end{equation}
In the above derivation, we have used the fact that the fields are
in the vacuum state in the initial state $|\Psi^{\prime}(0)\rangle$.
Note that $P_{j}^{{\rm sc}}(t)$ is just the semiclassical dynamics,
which is fully driven by the semiclassical Hamiltonian, that is,
\begin{equation}
\dot{U}_{{\rm S}}(t)=-iH_{{\rm S}}(t)U_{{\rm S}}(t).
\end{equation}
$P_{j}^{{\rm qc}}(t)$ represents quantum corrections to the semiclassical
dynamics, which is shown up to the second order in $g_{k}$. Clearly,
there are higher-order corrections, which can be computed similarly.
Such quantum corrections become vanishing in the limit of $g_{k}\rightarrow0$,
i.e., in the semiclassical limit.

Since in the semiclassical limit $|\alpha_k|\rightarrow\infty$, we calculate the change in the mean photon number of the mode
$k$, which is simply given by
\begin{eqnarray}
\Delta n_{k}(t) & = & \langle\Psi^{\prime}(t)|D^{\dagger}(\vec{\alpha})(b_{k}^{\dagger}b_{k}-|\alpha_{k}|^{2})D(\vec{\alpha})|\Psi^{\prime}(t)\rangle\nonumber \\
 & = & \langle\Psi^{\prime}(0)|U^{\prime\dagger}(t)(b_{k}^{\dagger}b_{k}+\alpha_{k}^{\ast}b_{k}+\alpha_{k}b_{k}^{\dagger})U^{\prime}(t)|\Psi^{\prime}(0)\rangle.\nonumber \\
\label{eq:Dnkt}
\end{eqnarray}
Plugging Eq. (\ref{eq:Upt}) into (\ref{eq:Dnkt}), using the fact
that the fields are in the vacuum state in the initial state $|\Psi^{\prime}(0)\rangle$
and $U_{{\rm S}}(t)$ commutes with the field operators, and taking
the semiclassical limit, we readily have
\begin{equation}
\Delta n_{k}(t)=-\Omega_{k}{\rm Re}\int_{0}^{t}i\langle V_{k}(\tau)\rangle_{0}e^{i(\omega_{k}\tau+\phi_{k})}d\tau.\label{eq:Dnktc}
\end{equation}
Similarly, we can calculate the variance in the photon number of the
mode $k$.
\begin{eqnarray}
\sigma_{k}^{2}(t) & = & \langle\Psi^{\prime}(t)|D^{\dagger}(\vec{\alpha})(b_{k}^{\dagger}b_{k}-|\alpha_{k}|^{2})^{2}D(\vec{\alpha})|\Psi^{\prime}(t)\rangle-\Delta n_{k}^{2}(t)\nonumber \\
 & = & |\alpha_{k}|^{2}+\Delta n_{k}(t)-\Delta n_{k}^{2}(t)\nonumber \\
 &  & +\frac{\Omega_{k}^{2}}{2}\int_{0}^{t}\int_{0}^{t}\langle V_{k}^{\dagger}(\tau_{1})V_{k}(\tau_{2})\rangle_{0}e^{-i\omega_{k}(\tau_{1}-\tau_{2})}d\tau_{1}d\tau_{2}\nonumber \\
 &  & -\Omega_{k}^{2}{\rm Re}\int_{0}^{t}d\tau_{1}\int_{0}^{\tau_{1}}d\tau_{2}\langle V_{k}(\tau_{1})V_{k}(\tau_{2})\rangle_{0}\nonumber \\
 &  & \times e^{i\omega_{k}(\tau_{1}+\tau_{2})+2i\phi_{k}}.\label{eq:varktc}
\end{eqnarray}
When $t=0$, one simply has $\sigma_{k}^{2}(0)=|\alpha_{k}|^{2}$,
which is the feature of a coherent state. In practice, since $|\alpha_{k}|^{2}$
is a large number in the semiclassical limit, we subtract $|\alpha_{k}|^{2}$
from $\sigma_{k}^{2}(t)$ and calculate the change in the variance,
i.e.,
\begin{equation}
\Delta\sigma_{k}^{2}(t)=\sigma_{k}^{2}(t)-|\alpha_{k}|^{2}. \label{eq:varph}
\end{equation}
This quantity characterizes the variation of the width of photon statistical distribution.
Note that Eqs.~(\ref{eq:Dnktc}) and (\ref{eq:varktc}) are only justified in the semiclassical limit $g_k\rightarrow 0$, $\alpha_k\rightarrow\infty$, and $g_k|\alpha_k|=\Omega_k$.
In other words, they are inapplicable in the quantum and crossover regimes. As long as the semiclassical time-evolution operator is obtained,
we can easily compute $\Delta n_{k}(t)$ and $\Delta\sigma_{k}^{2}(t)$.
The time-evolution operator $U_{{\rm S}}(t)$ can be computed via
directly integrating the time-dependent Schr\"{o}dinger equation.
Alternatively, it can be numerically or analytically computed with
(generalized) Floquet theory.

\section{Applications}\label{sec:app}

In this section, we apply the present variational approach and the
semiclassical approach to study the dynamics of some light-matter systems.
In doing so, we address the consistency between the quantum
variational dynamics and semiclassical dynamics in the presence of a large number of photons and explore the role of quantum corrections.
In the following calculations, the quantum system is assumed to be initially in the ground state while the field is initially in a coherent state.
Hereafter, the variational results will be denoted by ``$M$-$D_2$'' with $M$ specifying a concrete number of coherent states used in simulations.
The semiclassical results calculated from Eqs.~(\ref{eq:pjsc}), (\ref{eq:Dnktc}), and (\ref{eq:varph}) will be denoted by ``SC'' in the plots.

\subsection{Jaynes-Cummings model}

To begin with, we consider the  JC model, the Hamiltonian
of which reads
\begin{equation}
H=\frac{1}{2}\omega_{0}\sigma_{z}+\omega b^{\dagger}b+\frac{g}{2}(b\sigma_{+}+b^{\dagger}\sigma_{-}),
\end{equation}
where $\omega_{0}$ is the transition frequency between two levels:
$|1\rangle$ and $|2\rangle$, and $\sigma_{z}=|2\rangle\langle2|-|1\rangle\langle1|$
and $\sigma_{+}=\sigma_{-}^{\dagger}=|2\rangle\langle1|$. The notations for the field part are the same as in Eq.~(\ref{eq:Ham}) and since there
is a single mode, we have neglected the subscripts that label different
modes for the field. This model and its semiclassical counterpart are
exactly solvable, which provides transparent insights into the system
and field dynamics as well as the quantum corrections to the semiclassical
dynamics. For the quantized JC model, we use the variational approach
to calculate the dynamics of the system and field.
In the following, we introduce some analytical results from the semiclassical JC model.

\begin{figure*}
    \includegraphics[width=2\columnwidth]{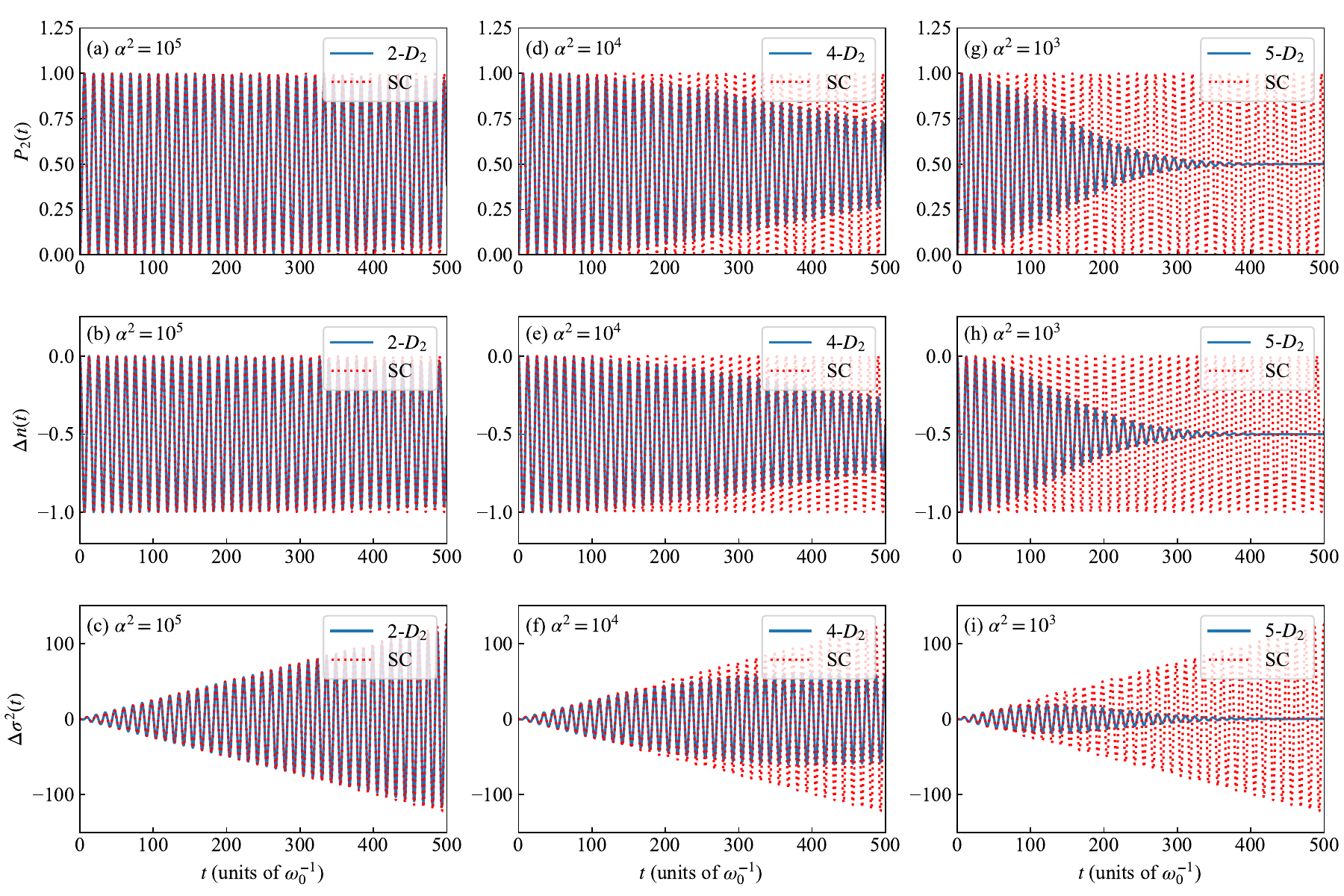}
    
    \caption{Excited-state population $P_{2}(t)$, change in photon number $\Delta n(t)$,
    and change in the variance of photon number $\Delta\sigma^{2}(t)$ calculated
    by the quantum variational approach and semiclassical approach for the quantum and semiclassical JC model. For the semiclassical model, the parameters are set as
    $\Omega=0.5\omega_{0}$ and $\omega=\omega_{0}$. For the quantum model, $g\alpha=\Omega$, $\omega=\omega_{0}$, and the three
    values of the initial mean photon number $\alpha^{2}$ are used. ``$2$-$D_2$'' denotes the variational results with $M=2$. ``SC'' denotes the semiclassical results.} \label{fig1}
    
\end{figure*}
    
\begin{figure}
        \includegraphics[width=1\columnwidth]{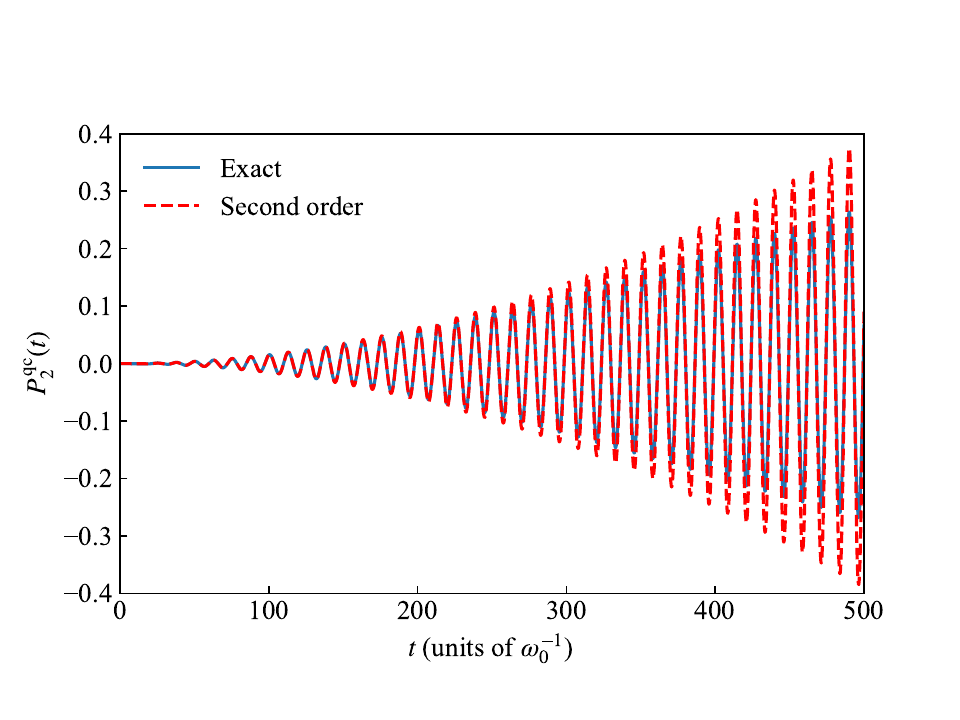}
        
        \caption{Exact and second-order quantum correction to the excited-state population of the qubit in the JC model for
        $\Omega=g\alpha=0.5\omega_{0}$, $\omega=\omega_{0}$, and $\alpha^{2}=10^4$. The exact quantum correction is obtained 
        as the difference between the quantum dynamics $P_2(t)$ and 
        the semiclassical dynamics $P_2^{\rm sc}(t)$. The second-order quantum correction is given by Eq.~(\ref{eq:p2qc}).} \label{fig2}
        
\end{figure}
    
\begin{figure*}
        \includegraphics[width=2\columnwidth]{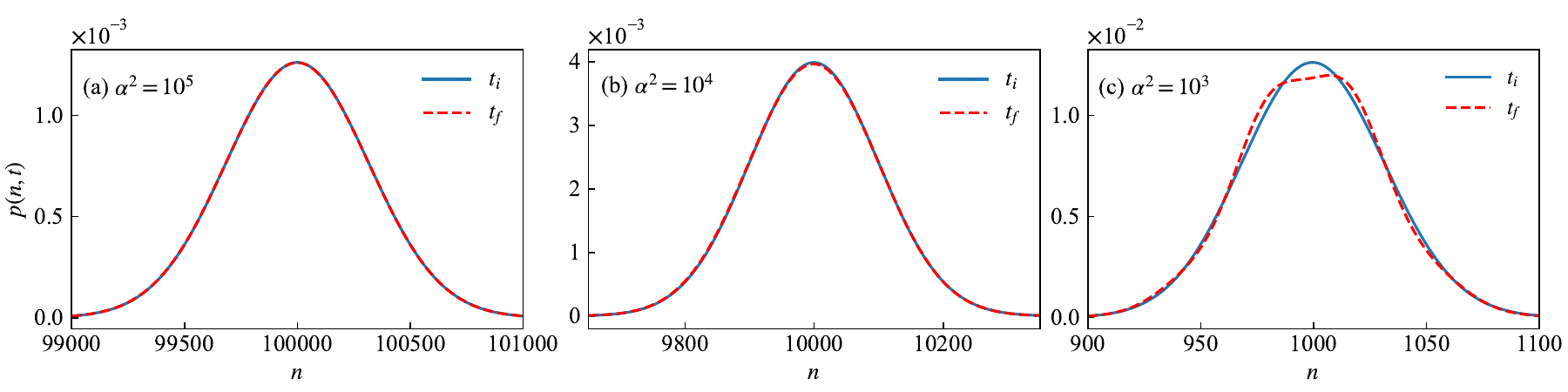}
        
        \caption{Photon-number distribution $p(n,t)$ as a function of photon number $n$ at the two given times calculated by the variational approach for 
        $\Omega=g\alpha=0.5\omega_{0}$, $\omega=\omega_{0}$, and the three values of $\alpha^2$. ``$t_i$'' and ``$t_f$'' denote the initial time ``$t=0$'' and finial time ``$t=500\omega_0^{-1}$'', respectively.} \label{fig3}
        
\end{figure*}

The semiclassical Hamiltonian of the JC model is given by
\begin{equation}
H_{{\rm S}}(t)=\frac{1}{2}\omega_{0}\sigma_{z}+\frac{\Omega}{2}(\sigma_{+}e^{-i\omega t}+\sigma_{-}e^{i\omega t}).
\end{equation}
The time-evolution operator reads~\cite{scully}
\begin{equation}
U_{{\rm S}}(t)  =  e^{-\frac{i\omega t\sigma_{z}}{2}}\left[\cos\left(\frac{\Omega_{{\rm R}}t}{2}\right)-i\sin\left(\frac{\Omega_{{\rm R}}t}{2}\right)\frac{\delta\sigma_{z}+\Omega\sigma_{x}}{\Omega_{{\rm R}}}\right], \label{eq:JCUst}
\end{equation}
where $\delta=\omega_{0}-\omega$ is the detuning and $\Omega_{{\rm R}}=\sqrt{\Omega^{2}+\delta^{2}}$
is the Rabi frequency. With the time-evolution operator and considering
the initial state of the two-level system $|\psi(0)\rangle=|1\rangle$,
we can easily obtain the excited-state population of the two-level
system
\begin{equation}
P_{2}^{{\rm sc}}(t)=\frac{\Omega^{2}}{\Omega_{{\rm R}}^{2}}\sin^{2}\left(\frac{\Omega_{{\rm R}}t}{2}\right).
\end{equation}
This is the celebrated Rabi oscillation. 

The change in photon number for the semiclassical JC model is obtained
as
\begin{equation}
\Delta n(t)=-P_{2}^{{\rm sc}}(t).
\end{equation}
This means that the change in photon number and the excited-state population of the two-level system
oscillate out of phase, which just reflects the conservation of the excitation number of the total system.

The change in the variance of the mean photon number is given by
\begin{eqnarray}
\Delta\sigma^{2}(t) & = & \frac{(\Omega^{2}-\delta^{2})\Omega^{2}}{\Omega_{{\rm R}}^{4}}\sin^{2}\left(\frac{\Omega_{{\rm R}}t}{2}\right)-\frac{\Omega^{4}}{\Omega_{{\rm R}}^{4}}\sin^{4}\left(\frac{\Omega_{{\rm R}}t}{2}\right)\nonumber \\
 &  & -\frac{\Omega^{4}t}{2\Omega_{{\rm R}}^{3}}\sin(\Omega_{{\rm R}}t).
\end{eqnarray}
The above analytical results on the field part are obtained in the
semiclassical limit and quantum corrections are not involved.

We now address how the deviation in the dynamics between the quantum and semiclassical JC model emerges due to the variation of the
the initial mean numbers of photons by comparing the variational and analytical results.
In Fig.~\ref{fig1}, we show the dynamics of the system and field
by computing the excited-state population $P_{2}(t)$, the change
in photon number $\Delta n(t)$, and the change in variance of photon number
$\Delta\sigma^{2}(t)$ from the quantum and semiclassical JC models for $\omega=\omega_{0}$. For the quantum model, we consider
the three values of the initial mean number of photons $\alpha^{2}$
ranging from $10^{5}$ to $10^{3}$. The
coupling constant and the Rabi frequency are set by $g\alpha=\Omega=0.5\omega_{0}$.
Figures~\ref{fig1}(a)-\ref{fig1}(c) show that when $\alpha^{2}=10^{5}$, the quantum and
semiclassical dynamics for either the system or field are in perfect agreement
in the time interval, indicating the consisitency between the quantum and semiclassical models in the large mean photon number limit. Figures~\ref{fig1}(d)-\ref{fig1}(f)
show that when $\alpha^{2}=10^{4}$, the quantum
variational dynamics and the semiclassical dynamics are almost the same when $t<200\omega_0^{-1}$ while deviate from each other when $t>200\omega_0^{-1}$. Moreover,
Figs.~\ref{fig1}(g)-\ref{fig1}(i) show that when $\alpha^{2}=10^{3}$,
the quantum variational dynamics just coincides with the
semiclassical dynamics in the first few cycles of Rabi oscillation.
As the time goes on, one readily notes that for full quantum dynamics,
the Rabi oscillation in the population of the system experiences
a significant collapse, which is a quantum feature and is absent in the semiclassical limit~\cite{scully,Everitt_2009}. The striking difference between
the quantum and the semiclassical results reflects the role of quantum
corrections to the semiclassical dynamics in the quantum-semiclassical
crossover.

To provide a criterion for assessing the time scale within which the quantum and semiclassical models yield consistent dynamics, we further explore the role of the quantum corrections. 
We calculate the difference between the variational population dynamics  $P_2(t)$
and the semiclassical dynamics $P_2^{\rm sc}(t)$, which is just the ``exact'' quantum correction. Alternatively, we can calculate the quantum correction up to the second order in $g$
for the JC model. It follows from Eqs.~(\ref{eq:qc}) and~(\ref{eq:JCUst}) that the second-order quantum correction to the semiclassical population dynamics is given by
\begin{eqnarray}
    P_{2}^{{\rm qc}}(t) & = & \frac{g^{2}\Omega^{2}}{4\Omega_{{\rm R}}^{4}}\left\{ \frac{\Omega^{2}t^{2}}{4}\cos(\Omega_{{\rm R}}t)+\frac{4\delta^{2}-\Omega^{2}}{4\Omega_{R}}t\sin(\Omega_{{\rm R}}t)\right.\nonumber\\
    &  &\left.-\frac{4\delta^{2}}{\Omega_{{\rm R}}^{2}}\sin^{2}\left(\frac{\Omega_{{\rm R}}t}{2}\right)\right\}. \label{eq:p2qc}
\end{eqnarray}
The analytical result shows that the amplitude
of oscillation is proportional to $t^{2}$. This means that even if $g$ is a small quantity, the quantum correction can contribute significantly to the long-time limit. Moreover,
this result is ill-defined as $t\rightarrow\infty$, suggesting that the second-order quantum correction to the semiclassical dynamics is insufficient 
in the quantum-semiclassical crossover and higher-order quantum corrections must be involved.
In Fig.~\ref{fig2}, we plot the numerically exact and second-order quantum corrections as a function of time $t$ for $\alpha^2=10^4$, $\Omega=g\alpha=0.5\omega_0$, and $\omega=\omega_0$.
One readily finds that the second-order quantum correction plays a predominant role and agrees with the exact one in the finite time interval.
This finding actually can be used to estimate the upper bound of time $t_c$ below which
the quantum dynamics and semiclassical dynamics are consistent. Roughly speaking, we require $g^2t_{c}^2<1$ such that the second-order correction is a small quantity.
Using $g=\Omega/|\alpha|$, we have
$t_c<|\alpha|/\Omega$, that is, the larger the mean photon number is, the more the quantum and semiclassical dynamics coincide with each other over a longer time interval. 
For instance, when considering the parameters in Fig.~\ref{fig1}(c), we find that $t_c\approx63\omega_{0}^{-1}$, which turns out to be a good estimation of the time scale.

The present variational approach can also be used to calculate the
photon-number distribution at given times. Figures~\ref{fig3}(a)-\ref{fig3}(c)
show $p(n,t)$ as a function of $n$ at two given times for $\omega=\omega_{0}$,
$g=0.5\omega_{0}/\alpha$, and three values of $\alpha$. When $t=0$, the photon-number distribution
is the Poisson distribution peaked at $n=|\alpha|^{2}$, which is
the nature of the coherent state. When $\alpha^2=10^{5}$ or $10^4$, we see that
there is almost no difference between the initial ($t=0$) and final ($t=500\omega_{0}^{-1}$) photon-number
distribution. This finding is consistent with the results in Figs.~\ref{fig1}(c) and~\ref{fig1}(f), where one finds that
although the oscillation amplitude of the change in the variance of photon number $\Delta\sigma^{2}(t)$ increases with time $t$, its magnitude is far
smaller than the initial variance $\sigma^{2}(0)=\alpha^2$, i.e., $\Delta\sigma^{2}(t)\ll\sigma^{2}(0)$. Consequently, the width of the photon-number distribution hardly changes in the semiclassical limit.
However, when $\alpha^2=10^3$, Fig.~\ref{fig3}(c) shows that the final photon-number distribution is different from the initial one. The present results suggest that the
photon-number distribution is rarely changed in the semiclassical
limit but can be changed due to the
light-matter interaction in the crossover region between the quantum and
semiclassical limit.

The present results suggest that the variational approach
is applicable to light-matter systems not only in the quantum limit~\cite{Werther_2018} but also in the semiclassical limit as well as the crossover in between, and is feasible to tackle
a relatively large mean number of photons. On the other
hand, the present results confirm that field dynamics can also
be calculated in the semiclassical model.

\subsection{Rabi model}

\subsubsection{Single-mode case}
The JC model may be inadequate due to the RWA and we move to consider the quantum Rabi model, the Hamiltonian of which reads
\begin{equation}
H=\frac{1}{2}\omega_{0}\sigma_{z}+\omega b^{\dagger}b+\frac{g}{2}(b+b^{\dagger})\sigma_{x},
\end{equation}
where $\sigma_{x}=\sigma_{+}+\sigma_{-}$.
We apply the variational approach to study the dynamics of the
quantum Rabi model. We also calculate the semiclassical dynamics of
the system and field, which is based on the numerically calculated
time-evolution operator for the semiclassical Rabi model: $H_{\rm S}(t)=\frac{1}{2}\omega_0\sigma_{z}+\Omega\cos(\omega t)\sigma_x$.

\begin{figure*}
    \includegraphics[width=2\columnwidth]{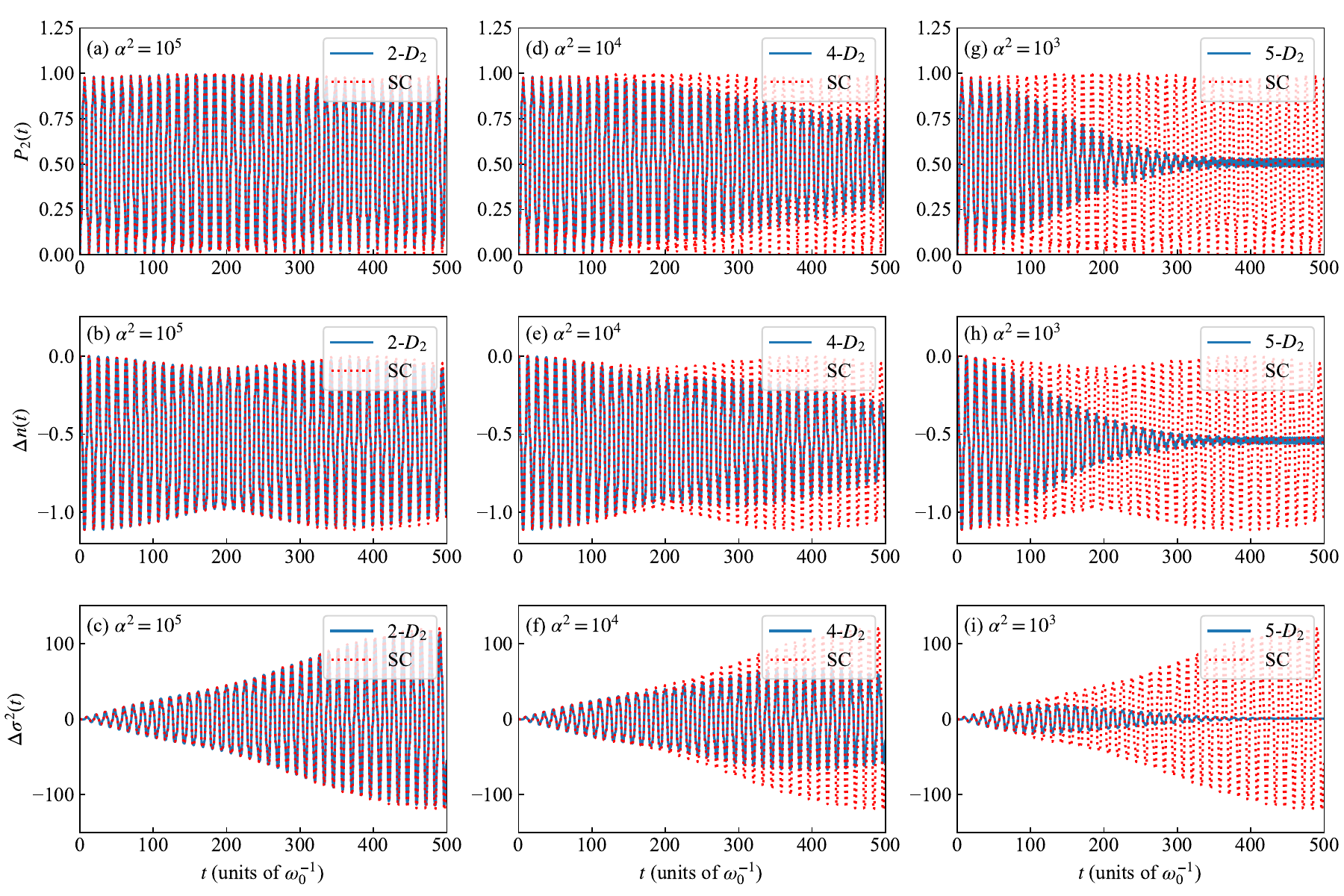}
    
    \caption{Excited-state population $P_{2}(t)$, change in photon number $\Delta n(t)$,
    and change in the variance of photon number $\Delta\sigma^{2}(t)$ calculated
    by the quantum variational approach and semiclassical approach for the quantum and semiclassical Rabi model. For the semiclassical model, the parameters are set as
    $\Omega=0.5\omega_{0}$ and $\omega=\omega_{0}$. For the quantum model, $g\alpha=\Omega$, $\omega=\omega_{0}$, and the three
    values of the initial mean photon number $\alpha^{2}$ are used.} \label{fig4}
    
\end{figure*}

We examine the consistency between the variational dynamics and the semiclassical dynamics under the initial condition of a large number of photons, i.e., $\alpha^2\gg1$.
In Fig.~\ref{fig4}, we show the dynamics of the excited-state population $P_2(t)$, the change in photon number $\Delta n(t)$, and
the change in the variance of the mean photon number $\Delta\sigma^{2}(t)$
for the quantum and semiclassical Rabi model. We consider the three values of $\alpha^2$ and $\omega=\omega_0$. The Rabi frequency and coupling constant are given by
$\Omega=0.5\omega_0$ and $g=\Omega/\alpha$, respectively. One notes that the quantum variational dynamics and semiclassical
dynamics are almost the same when $\alpha^2=10^5$. However, when $\alpha^2=10^4$ or $10^3$, the difference between the
variational and semiclassical dynamics appears as the time increases, which is attributed to the quantum corrections to the semiclassical dynamics.
This situation is similar to that encountered in the JC model. In addition, one notes that there are beat behaviors in Fig.~\ref{fig4}, which results from
the counter-rotating terms~\cite{Shirley_1965,Stenholm_1972,Zhiguo_2012}.

\begin{figure*}
    \includegraphics[width=2\columnwidth]{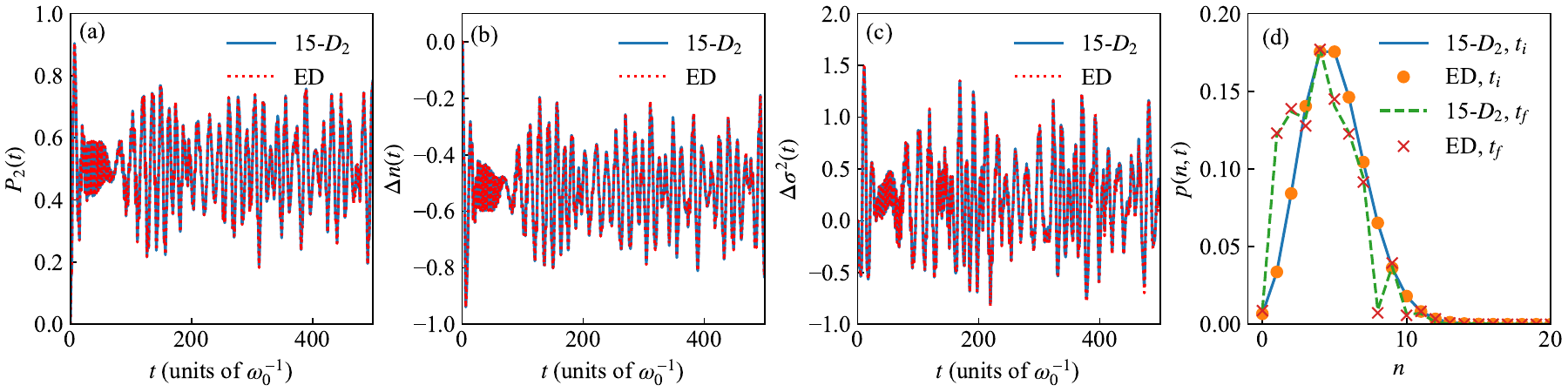}
    
    \caption{(a) Excited-state population $P_{2}(t)$, (b) change in photon number $\Delta n(t)$,
    (c) change in the variance of photon number $\Delta\sigma^{2}(t)$, and (d)  photon-number distribution $p(n,t)$ calculated
    by the variational approach and numerically exact diagonalization (ED) for the quantum Rabi model.
    The parameters are set as
    $g=0.2\omega_{0}$, $\omega=\omega_{0}$, and $\alpha^2=5$. ``$t_i$'' and ``$t_f$'' in panel (d) denote the initial time ``$t=0$'' and finial time ``$t=500\omega_0^{-1}$'', respectively.}\label{fig5}
    
\end{figure*}

To explore the validity of the variational approach in the quantum
limit, we calculate the dynamics of the system and field for the quantum
Rabi model by using the variational approach and the numerically exact diagonalization (ED)
of the quantum Rabi Hamiltonian for $\omega=\omega_{0}$, $g=0.2\omega_{0}$,
and $\alpha^{2}=5$. Since $g=0.2\omega_{0}$ is comparable to $\omega$,
the light-matter coupling is ultrastrong. This is a regime where the two-level system ultrastrongly interacts with a few photons, and thus the semiclassical
approach is inapplicable. Figure~\ref{fig5}
shows that in comparison with the ED method, the variational approach can produce the accurate dynamics
of system and field as well as the photon-number distribution for the quantum Rabi model in an ultrastrong coupling regime. The present findings
suggest that the variational approach provides a unified method to
tackle light-matter systems in both the quantum and semiclassical
limits. More importantly, it captures the dynamics of both system and field. This is of particular importance in
potential applications of studying statistical properties of the field.

\subsubsection{Two-mode case}

We now exploit the developed formalisms to study the system and field dynamics in the two-mode cases,
where multiphoton processes become remarkable. The Hamiltonian of the two-mode quantum Rabi model is given by
\begin{equation}
H=\frac{\omega_{0}}{2}\sigma_{z}+\sum_{k=1}^{N}\omega_{k}b_{k}^{\dagger}b_{k}+\sum_{k=1}^{N}\frac{g_{k}}{2}(b_{k}+b_{k}^{\dagger})\sigma_{x},
\end{equation}
where the number of modes is $N=2$. The Hamiltonian of the two-mode semiclassical Rabi model
is given by
\begin{equation}
 H_{\rm S}(t)=\frac{\omega_{0}}{2}\sigma_{z}+[\Omega_1\cos(\omega_1 t)+\Omega_{2}\cos(\omega_2 t)]\sigma_{x},
\end{equation}
which describes a bichromatically driven two-level system. As is well-known, the two-level system may absorb $n+1$ photons from one field and emit $n$ photons into the other field,
which can lead to the multiphoton resonances occurring at $\omega_0\approx (n+1)\omega_1-n\omega_2$ with $n$ an integer. Such multiphoton resonances
have been studied in the semiclassical model with and without the RWA~\cite{Guccione_Gush_1974,Ho_1984,Ruyten_1989,Ruyten_1992,Yan_2023}, and are illustrated by calculating the transition probabilities of the two-level system. Here we revisit this phenomenon by
calculating not only the population of the system but also the photon-number dynamics by making use of variational and semiclassical approaches. 
In the latter approach, the time-evolution operator of the semiclassical model is numerically calculated by the Runge-Kutta method.

\begin{figure*}
    \includegraphics[width=2\columnwidth]{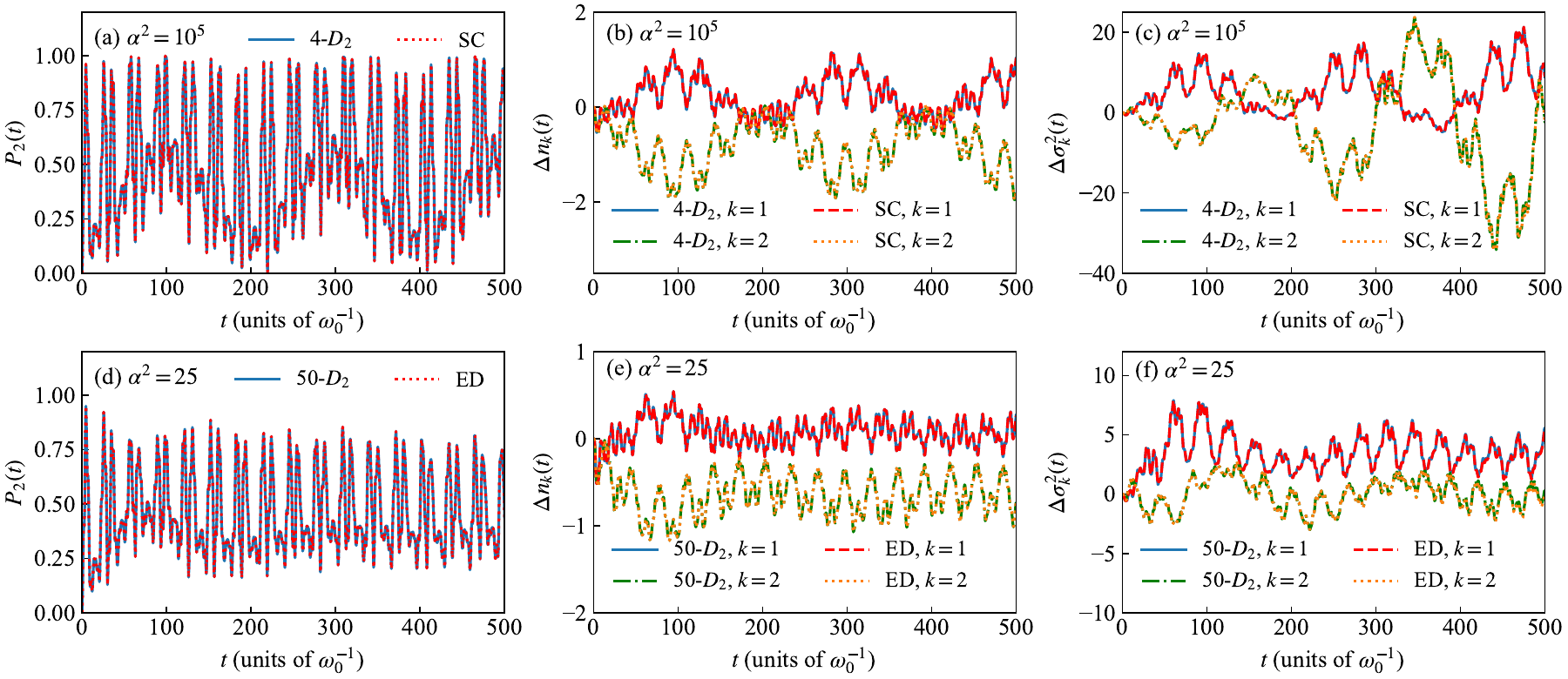}
    
    \caption{Excited-state population $P_{2}(t)$, change in photon number $\Delta n_{k}(t)$,
    and change in the variance of photon number $\Delta\sigma^{2}_{k}(t)$ calculated
    by the variational approach, semiclassical approach [(a)-(c)], and the ED method [(d)-(f)] for the two-mode quantum and semiclassical Rabi model. For the semiclassical model, the parameters are set as
    $\omega_{1}=0.6449\omega_{0}$, $\omega_2=0.8449\omega_{0}$, and $\Omega_{1}=\Omega_2=0.3\omega_0$. 
    For the quantum model, $\omega_{1}=0.6449\omega_{0}$, $\omega_2=0.8449\omega_{0}$, $g_1=g_2=\Omega_{1}/\alpha$, and $\alpha_{1}=\alpha_{2}=\alpha$. The two
    values of the initial mean photon number $\alpha^{2}$ are used.} \label{fig6}
    
\end{figure*}

\begin{figure*}
    \includegraphics[width=\columnwidth]{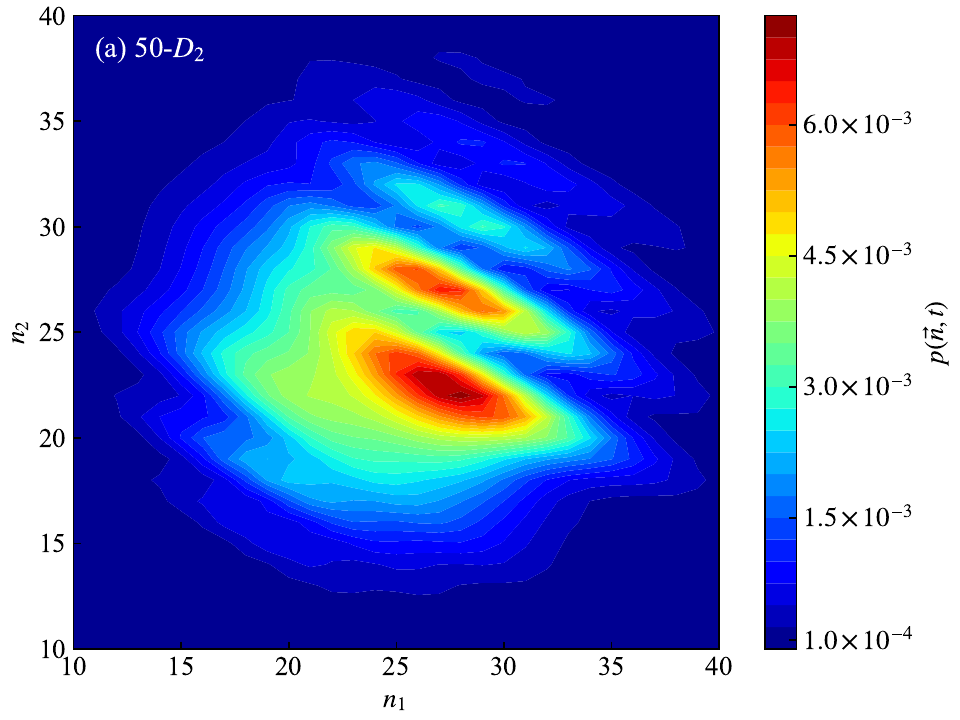}
    \includegraphics[width=\columnwidth]{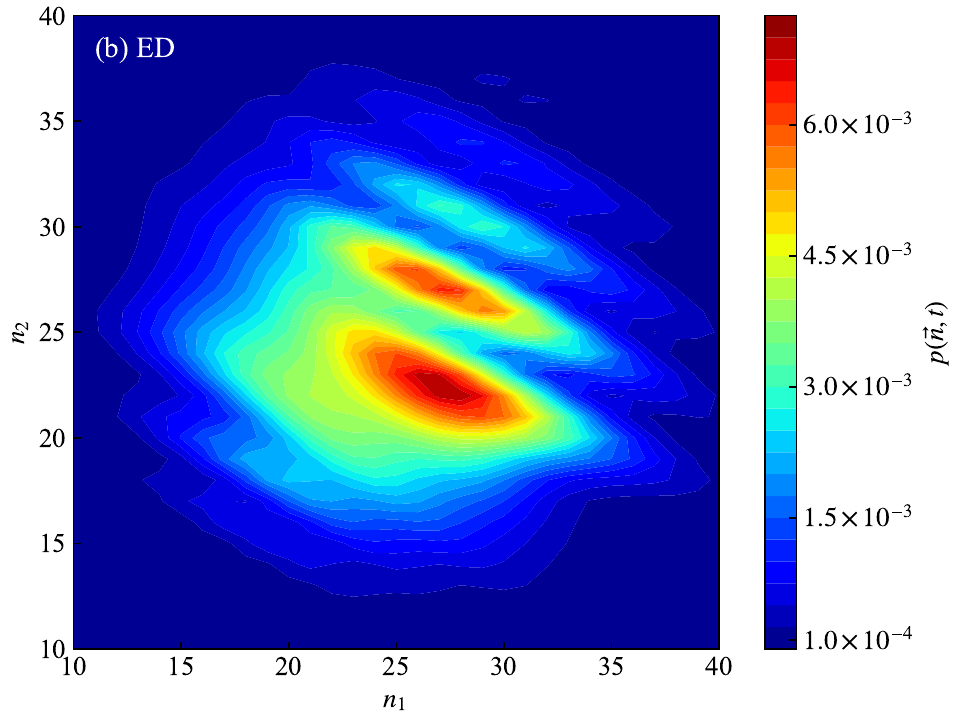}
    \caption{Photon-number distribution $p(\vec{n},t)$ with $\vec{n}=(n_1,n_2)$ at $t=500\omega_0^{-1}$ for the two-mode quantum Rabi model calculated by the variational approach (a) and the ED method  (b).
    The parameters are the same as in Fig.~\ref{fig6}(d).}\label{fig7}
    
\end{figure*}

The multiphoton resonance condition can be numerically obtained by the generalized Floquet theory~\cite{Ho_1984}.
To be concrete, we consider $\omega_1=0.6449\omega_0$, $\omega_2=0.8449\omega_0$, and $\Omega_1=\Omega_{2}=0.3\omega_0$ for the semiclassical model, which corresponds to
the resonance that the two-level system absorbs two photons from the mode 2 and emit one photon into the mode 1.
For the quantized field, we set $g_1=g_2=0.3\omega_0/\alpha$ and $\alpha_1^2=\alpha_{2}^2=\alpha^2=10^5$.  In Figs.~\ref{fig6}(a)-\ref{fig6}(c), we depict the excited-state
population $P_2(t)$, the change in photon number $\Delta n_{k}(t)$, and the change in the variance $\Delta\sigma^2_k(t)$ for the quantum and semiclassical models.
It is evident that the dynamics of the system and fields from the quantum model are in agreement with
those from the semiclassical model, confirming that the variational approach can be applied
to simulate the semiclassical dynamics in the two-mode cases provided the initial mean photon number
is sufficiently large. In Fig.~\ref{fig6}(a), we see that the excited-state population can reach a maximum value $P_2(t)=1$, which signifies the occurrence of resonance. In Fig.~\ref{fig6}(b),
the change in photon numbers in the two modes manifests the multiphoton feature of the resonance.

Figures~\ref{fig6}(d)-\ref{fig6}(f) show the dynamics of the system and fields for the model with quantized fields
in the case of $\alpha^2_{1}=\alpha^2_{2}=\alpha^2=25$, $g_1=g_2=\Omega_{1}/\alpha=0.06\omega_0$. Comparisons between the variational results and those of the ED method confirm that
the variational approach is applicable in the quantum limit in the two-mode case. In addition, we see that
the system and field dynamics in the case of $\alpha^2=25$ are apparently different from those
in the case of $\alpha^2=10^5$. Specifically,  the excited-state population cannot reach a maximum value $P_2(t)=1$ in the former case, indicating the disappearance of
the multiphoton resonance due to the quantum corrections. In Figs.~\ref{fig7}(a)-\ref{fig7}(b), we compare the photon-number distribution
$p(\vec{n},t)$ with $\vec{n}=(n_1,n_2)$ at $t=500\omega_0^{-1}$ calculated from the variational and the ED methods for the same parameters as in Figs.~\ref{fig6}(d)-\ref{fig6}(f).
The two methods predict almost the same photon-number distribution, which is apparently different from the initial two-dimensional Poisson distribution due to the strong light-matter coupling.

\begin{figure*}
    \includegraphics[width=2\columnwidth]{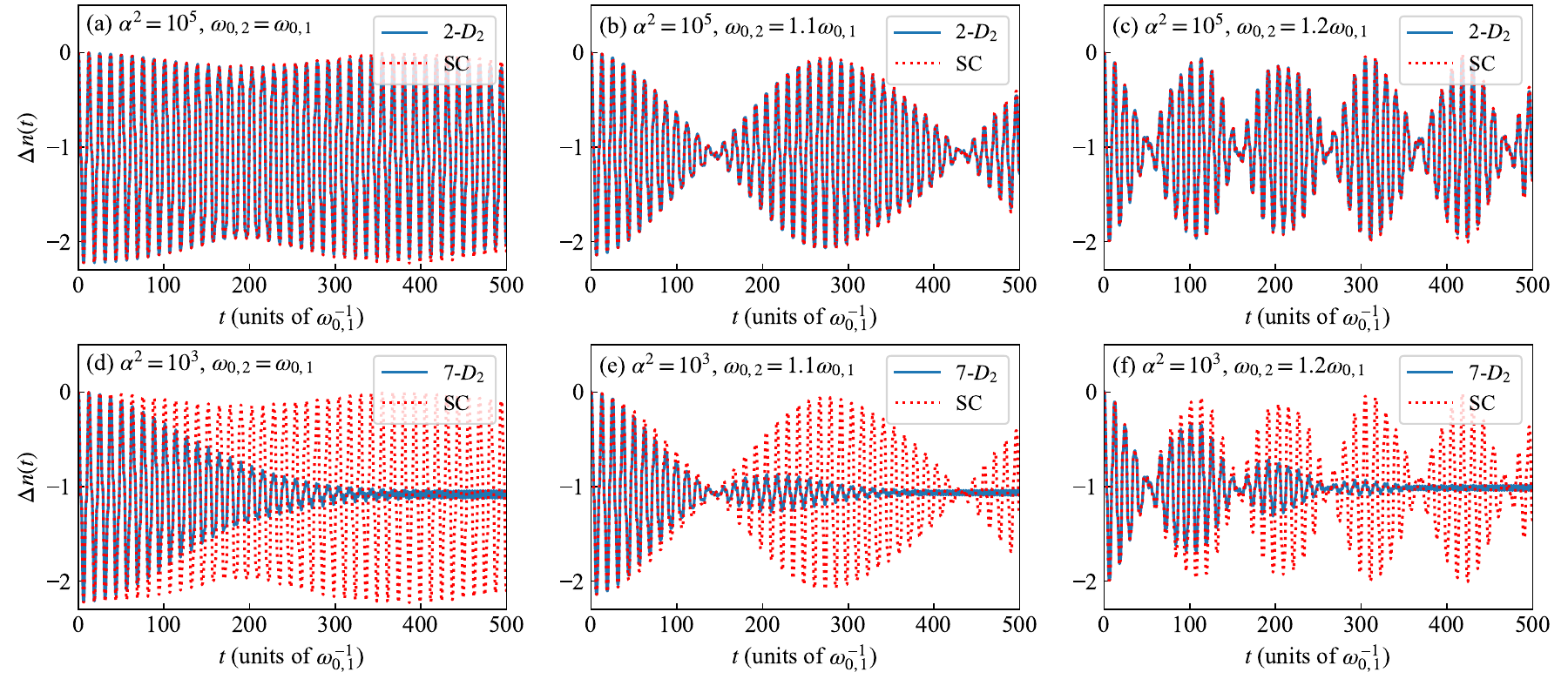}
    
    \caption{Change in photon number $\Delta n(t)$ calculated
    by the variational approach and semiclassical approach for the two-qubit Dicke model and its semiclassical counterpart with $N_q=2$.
    For the semiclassical model, $\Omega=0.5\omega_{0,1}$, $\omega=\omega_{0,1}$, and the three values of $\omega_{0,2}$ are used. For the quantum model, $g\alpha=\Omega$, 
    the two values of initial mean photon number $\alpha^{2}$ are used, and the other parameters are the same as in the semiclassical model.}\label{fig8}
    
\end{figure*}

\subsection{Dicke model}
In this section, we consider the Dicke model~\cite{Dicke_1954}, which is described by the following Hamiltonian
\begin{equation}
    H=\omega b^\dagger b+\frac{1}{2}\sum_{j=1}^{N_{q}}[\omega_{0,j}\sigma_{z,j}+g(b+b^\dagger)\sigma_{x,j}],
\end{equation}
where $N_{q}$ is the number of the two-level systems, $\omega_{0,j}$ is the transition frequency of the $j$th two-level system,
$\sigma_{\mu,j}$ is the Pauli matrix for the $j$th two-level system.
The semiclassical counterpart of the Dicke model is given by
\begin{equation}
    H_{\rm S}(t)=\sum_{j=1}^{N_{q}}\left[\frac{1}{2}\omega_{0,j}\sigma_{z,j}+\Omega\cos(\omega t)\sigma_{x,j}\right].
\end{equation}

Figure~\ref{fig8} shows the change in photon number $\Delta n(t)$ calculated by the variational and semiclassical approaches for $N_q=2$, $\omega=\omega_{0,1}$,
$\Omega=0.5\omega_{0,1}$, and the three values of $\omega_{0,2}$. For the quantum model, we consider two values of $\alpha^2$ and $g=\Omega/\alpha$. 
In Figs.~\ref{fig8}(a)-\ref{fig8}(c), 
we see that when $\alpha=10^5$, the variational quantum dynamics coincides with the semiclassical dynamics. Figures~\ref{fig8}(d)-\ref{fig8}(f) show that when $\alpha^2=10^3$, the oscillation from the quantum model 
undergoes collapse. The present results on the consistency between the quantum and semiclassical models
are similar to that in the JC and Rabi models.

Figure~\ref{fig8} also provides insights into how the field responds due to the presence of multiple two-level systems.
When the two-level systems are identical, i.e., $\omega_{0,2}=\omega_{0,1}$, the photon number dynamics is similar to that of the Rabi model shown in Fig.~\ref{fig4}(b) except that the amplitude of the oscillation
is about 2, which reflects the fact that the two photons can be absorbed by the two subsystems.
On the other hand, when the two-level systems have different transition frequencies, there are strong beat behaviors that can be tuned by the frequency difference of the two subsystems, which result from the
two independent Rabi oscillations of the two subsystems that have different Rabi frequencies.

\section{Conclusions}\label{sec:con}

In summary, we have presented a time-dependent variational approach to study the system and field dynamics of
light-matter systems when the field is in a coherent state and possesses a finite mean number of photons.
In addition to the variational approach, we have shown that the field dynamics can also be deduced
from the system dynamics in the semiclassical model. By using the variational and semiclassical approaches, we have examined
the consistency in the system and field dynamics between the quantum and semiclassical models of light-matter interaction in the large mean photon number regimes.
We have illustrated that the variational approach can produce accurate
semiclassical dynamics of either the system or the field as long as the initial mean photon number is sufficiently large. Moreover, it can also produce accurate quantum dynamics of the system and field when
a few photons strongly interact with the quantum system and can apply to the crossover between the quantum and
semiclassical limits. In the crossover region, we have shown that the excited-state population of the quantum system and the change in photon number
experiences a collapse in the amplitude of the oscillations, which reflects the quantum feature.
The present variational approach provides a unified treatment of light-matter interaction in the quantum and semiclassical limits as well as the crossover in between.

The variational approach can also be extended to open quantum systems based on two routines. 
One is that the dissipation is taken into account by considering a set of harmonic oscillators to be the environment 
such as the well-known spin-boson model~\cite{Zhao_2023,Zueco_2019}. The other is that the dissipation is modeled by non-Hermitian Hamiltonians. 
The variational approach with the Davydov ansatz can be extended to solve the time-dependent Schr\"{o}dinger equation with a non-Hermitian Hamiltonian. 
This has the potential to describe cavity-QED systems with a lossy cavity.

\begin{acknowledgments}
    Support from the National Natural Science Foundation of China (Grant No.~12005188, No.~11774226, and No.~11774311)
    is gratefully acknowledged.
\end{acknowledgments}

\appendix*
\section{The equations of motion for the variational parameters and numerical implementation}
The variation of the joint state of $|D_{2}^{M}(t)\rangle$ is given
by
\begin{eqnarray}
\langle\delta D_{2}^{M}(t)|& = &\sum_{l=1}^{M}\sum_{j=1}^{N_{S}}\langle j|\langle f_{l}|\left\{ \delta A_{lj}^{\ast}+A_{lj}^{\ast}\sum_{k=1}^{N}\left[\delta f_{lk}^{\ast}\right.\right.\nonumber\\
&  &\times\left.\left.\left(b_{k}-\frac{1}{2}f_{lk}\right)-\frac{1}{2}\delta f_{lk}f_{lk}^{*}\right]\right\}.
\end{eqnarray}
Substituting $\langle\delta D_{2}^{M}(t)|$ into Eq.~(\ref{eq:tdvp}), one simply
derives
\begin{eqnarray}
0 & = & \sum_{l=1}^{M}\sum_{j=1}^{N_{S}}\delta A_{lj}^{\ast}\langle j|\langle f_{l}|\left[i\partial_{t}-H^{\prime}(t)\right]|D_{2}^{M}(t)\rangle\nonumber \\
 &  & +\sum_{l=1}^{M}\sum_{k=1}^{N}\delta f_{lk}^{\ast}\sum_{j=1}^{N_{S}}A_{lj}^{\ast}\langle j|\langle f_{l}|\left(b_{k}-\frac{1}{2}f_{lk}\right)\nonumber\\
 &   &\times\left[i\partial_{t}-H^{\prime}(t)\right]|D_{2}^{M}(t)\rangle\nonumber \\
 &  & -\sum_{l=1}^{M}\sum_{j=1}^{N_{S}}A_{lj}^{\ast}\sum_{k=1}^{N}\frac{\delta f_{lk}f_{lk}^{*}}{2}\langle j|\langle f_{l}|\left[i\partial_{t}-H^{\prime}(t)\right]|D_{2}^{M}(t)\rangle.\nonumber\\
\end{eqnarray}
To ensure the above equation holds for arbitrary variation of the parameters,
we simply have
\begin{equation}
\langle j|\langle f_{l}|\left[i\partial_{t}-H^{\prime}(t)\right]|D_{2}^{M}(t)\rangle=0,\label{eq:appeq1}
\end{equation}
\begin{equation}
\sum_{j=1}^{N_{S}}A_{lj}^{\ast}\langle j|\langle f_{l}|\left(b_{k}-\frac{1}{2}f_{lk}\right)\left[i\partial_{t}-H^{\prime}(t)\right]|D_{2}^{M}(t)\rangle=0.\label{eq:appeq2}
\end{equation}
Equation (\ref{eq:appeq1}) corresponds to Eq.~(\ref{eq:eom1}) in the main text and
can be used to simplify Eq. (\ref{eq:appeq2}), which leads to Eq.~(\ref{eq:eom2})
in the main text.

To derive the explicit forms of the equations of motion, we use the
time derivative of the trial state~\cite{Werther_2020}
\begin{eqnarray}
|\dot{D}_{2}^{M}(t)\rangle & = & \sum_{n=1}^{M}\sum_{i=1}^{N_{S}}\left[a_{ni}+A_{ni}\sum_{p=1}^{N}\dot{f}_{np}b_{p}^{\dagger}\right]|i\rangle|f_{n}\rangle,
\end{eqnarray}
where
\begin{equation}
a_{ni}=\dot{A}_{ni}-\frac{1}{2}A_{ni}\sum_{p=1}^{N}(\dot{f}_{np}f_{np}^{\ast}+f_{np}\dot{f}_{np}^{\ast}).\label{eq:anj}
\end{equation}

To proceed, we carry out calculation for $\langle j|\langle f_{l}|\dot{D}_{2}^{M}(t)\rangle$,
$\sum_{j=1}^{N_{S}}A_{lj}^{\ast}\langle j|\langle f_{l}|b_{k}|\dot{D}_{2}^{M}(t)\rangle$,
$\langle j|\langle f_{l}|H^{\prime}(t)|D_{2}^{M}(t)\rangle$, and
$\sum_{j=1}^{N_{S}}A_{lj}^{\ast}\langle j|\langle f_{l}|b_{k}H^{\prime}(t)|D_{2}^{M}(t)\rangle$
to express them in terms of the variational parameters, which yields
\begin{widetext}
\begin{equation}
\langle j|\langle f_{l}|\dot{D}_{2}^{M}(t)\rangle=\sum_{n=1}^{M}\sum_{i=1}^{N_{S}}\left(a_{nj}+A_{nj}\sum_{p=1}^{N}f_{lp}^{\ast}\dot{f}_{np}\right){\cal S}_{ln},
\end{equation}
\begin{equation}
\sum_{j=1}^{N_{S}}A_{lj}^{\ast}\langle j|\langle f_{l}|b_{k}|\dot{D}_{2}^{M}(t)\rangle=\sum_{n=1}^{M}\sum_{j,i=1}^{N_{S}}\left(A_{lj}^{\ast}a_{nj}f_{nk}+A_{lj}^{\ast}A_{nj}\sum_{p=1}^{N}\left(\delta_{k,p}+f_{lp}^{\ast}f_{nk}\right)\dot{f}_{np}\right){\cal S}_{ln},
\end{equation}
\begin{eqnarray}
[\vec{I}_{j}]_{l} & = & \langle j|\langle f_{l}|H^{\prime}(t)|D_{2}^{M}(t)\rangle\nonumber \\
 & = & \sum_{n=1}^{M}\sum_{i=1}^{N_{S}}\langle j|H_{S}(t)|i\rangle A_{ni}{\cal S}_{ln}+\sum_{n=1}^{M}\sum_{i=1}^{N_{S}}\sum_{p=1}^{N}\frac{g_{p}}{2}A_{ni}(\langle j|V_{p}^{\dagger}|i\rangle e^{-i\omega_{p}t}f_{np}+\langle j|V_{p}|i\rangle e^{i\omega_{p}t}f_{lp}^{\ast}){\cal S}_{ln},
\end{eqnarray}
\begin{eqnarray}
[\vec{I}_{f}]_{lk} & = & \sum_{j=1}^{N_{S}}A_{lj}^{\ast}\langle j|\langle f_{l}|b_{k}H^{\prime}(t)|D_{2}^{M}(t)\rangle\nonumber \\
 & = & \sum_{n=1}^{M}\sum_{j,i=1}^{N_{S}}A_{lj}^{\ast}\langle j|H_{S}(t)|i\rangle A_{ni}{\cal S}_{ln}f_{nk}+\sum_{n=1}^{M}\sum_{j,i=1}^{N_{S}}\sum_{p=1}^{N}\frac{g_{p}}{2}A_{lj}^{\ast}A_{ni}[\langle j|V_{p}^{\dagger}|i\rangle e^{-i\omega_{p}t}f_{np}f_{nk}\nonumber \\
 &  & +\langle j|V_{p}|i\rangle e^{i\omega_{p}t}(\delta_{p,k}+f_{lp}^{\ast}f_{nk})]{\cal S}_{ln},
\end{eqnarray}
\end{widetext}
where $[\vec{I}_{f}]_{lk}$ is viewed as a column vector with its
component position being specified by $l$ and $k$. By substituting
these quantities into Eqs.~(\ref{eq:eom1}) and~(\ref{eq:eom2}), we can rewrite the equations of motion
in a matrix form
\begin{equation}
i\left(\begin{array}{ccccc}
{\cal S} &  &  &  & {\cal C}^{(1)}\\
 & {\cal S} &  &  & {\cal C}^{(2)}\\
 &  & \ddots &  & \vdots\\
 &  &  & {\cal S} & {\cal C}^{(N_{S})}\\
{\cal C}^{(1)\dagger} & {\cal C}^{(2)\dagger} & \cdots & {\cal C}^{(N_{S})\dagger} & {\cal D}
\end{array}\right)\left(\begin{array}{c}
\vec{a}_{1}\\
\vec{a_{2}}\\
\vdots\\
\vec{a}_{N_{S}}\\
\vec{\dot{F}}
\end{array}\right)=\left(\begin{array}{c}
\vec{I}_{1}\\
\vec{I}_{2}\\
\vdots\\
\vec{I}_{N_{S}}\\
\vec{I}_{f}
\end{array}\right),\label{eq:mateq}
\end{equation}
where ${\cal S}$ is an $M\times M$ matrix whose elements are ${\cal S}_{ln}$, $\vec{a}_{j}=(a_{1j},a_{2j},\ldots,a_{Mj})^{T},$ $\vec{\dot{F}}$
is a vector whose components are given by $\dot{f}_{np}$, ${\cal C}^{(j)}$
is an $M\times MN$ matrix whose elements are given by
\begin{equation}
{\cal C}_{l,np}^{(j)}=A_{nj}{\cal S}_{ln}f_{lp}^{\ast},
\end{equation}
and ${\cal D}$ is an $MN\times MN$ matrix whose matrix elements
are
\begin{equation}
{\cal D}_{lk,np}=\sum_{j=1}^{N_{S}}A_{lj}^{\ast}A_{nj}\left(\delta_{k,p}+f_{lp}^{\ast}f_{nk}\right){\cal S}_{ln}.
\end{equation}

To perform numerical simulation, we first numerically solve the matrix equation
(\ref{eq:mateq}) as a set of linear equations to obtain the values
of $\vec{a}_{j}$ and $\vec{\dot{F}}$. The former can be combined with Eq. (\ref{eq:anj})
to calculate the derivatives of $A_{nj}$. The latter are just the derivatives
of $f_{np}$.  On obtaining the derivatives
of $A_{nj}$ and $f_{np}$, we can use the 4th-order Runge-Kutta algorithm
to update the variational parameters.

\bibliography{field_dynamics}

\begin{thebibliography}{52}%
\makeatletter
\providecommand \@ifxundefined [1]{%
 \@ifx{#1\undefined}
}%
\providecommand \@ifnum [1]{%
 \ifnum #1\expandafter \@firstoftwo
 \else \expandafter \@secondoftwo
 \fi
}%
\providecommand \@ifx [1]{%
 \ifx #1\expandafter \@firstoftwo
 \else \expandafter \@secondoftwo
 \fi
}%
\providecommand \natexlab [1]{#1}%
\providecommand \enquote  [1]{``#1''}%
\providecommand \bibnamefont  [1]{#1}%
\providecommand \bibfnamefont [1]{#1}%
\providecommand \citenamefont [1]{#1}%
\providecommand \href@noop [0]{\@secondoftwo}%
\providecommand \href [0]{\begingroup \@sanitize@url \@href}%
\providecommand \@href[1]{\@@startlink{#1}\@@href}%
\providecommand \@@href[1]{\endgroup#1\@@endlink}%
\providecommand \@sanitize@url [0]{\catcode `\\12\catcode `\$12\catcode `\&12\catcode `\#12\catcode `\^12\catcode `\_12\catcode `\%12\relax}%
\providecommand \@@startlink[1]{}%
\providecommand \@@endlink[0]{}%
\providecommand \url  [0]{\begingroup\@sanitize@url \@url }%
\providecommand \@url [1]{\endgroup\@href {#1}{\urlprefix }}%
\providecommand \urlprefix  [0]{URL }%
\providecommand \Eprint [0]{\href }%
\providecommand \doibase [0]{https://doi.org/}%
\providecommand \selectlanguage [0]{\@gobble}%
\providecommand \bibinfo  [0]{\@secondoftwo}%
\providecommand \bibfield  [0]{\@secondoftwo}%
\providecommand \translation [1]{[#1]}%
\providecommand \BibitemOpen [0]{}%
\providecommand \bibitemStop [0]{}%
\providecommand \bibitemNoStop [0]{.\EOS\space}%
\providecommand \EOS [0]{\spacefactor3000\relax}%
\providecommand \BibitemShut  [1]{\csname bibitem#1\endcsname}%
\let\auto@bib@innerbib\@empty
\bibitem [{\citenamefont {Scully}\ and\ \citenamefont {Zubairy}(1997)}]{scully}%
  \BibitemOpen
  \bibfield  {author} {\bibinfo {author} {\bibfnamefont {M.~O.}\ \bibnamefont {Scully}}\ and\ \bibinfo {author} {\bibfnamefont {M.~S.}\ \bibnamefont {Zubairy}},\ }\href@noop {} {\emph {\bibinfo {title} {Quantum Optics}}}\ (\bibinfo  {publisher} {Cambridge University Press},\ \bibinfo {address} {Cambridge},\ \bibinfo {year} {1997})\BibitemShut {NoStop}%
\bibitem [{\citenamefont {Cohen-Tannoudji}\ \emph {et~al.}(2004)\citenamefont {Cohen-Tannoudji}, \citenamefont {Dupont-Roc},\ and\ \citenamefont {Grynberg}}]{cohen}%
  \BibitemOpen
  \bibfield  {author} {\bibinfo {author} {\bibfnamefont {C.}~\bibnamefont {Cohen-Tannoudji}}, \bibinfo {author} {\bibfnamefont {J.}~\bibnamefont {Dupont-Roc}},\ and\ \bibinfo {author} {\bibfnamefont {G.}~\bibnamefont {Grynberg}},\ }\href@noop {} {\emph {\bibinfo {title} {Atom-Photon Interactions: Basic Processes and Applications}}}\ (\bibinfo  {publisher} {Wiley-VCH},\ \bibinfo {address} {Weinheim},\ \bibinfo {year} {2004})\BibitemShut {NoStop}%
\bibitem [{\citenamefont {Rivera}\ and\ \citenamefont {Kaminer}(2020)}]{Rivera_2020}%
  \BibitemOpen
  \bibfield  {author} {\bibinfo {author} {\bibfnamefont {N.}~\bibnamefont {Rivera}}\ and\ \bibinfo {author} {\bibfnamefont {I.}~\bibnamefont {Kaminer}},\ }\bibfield  {title} {\bibinfo {title} {Light-matter interactions with photonic quasiparticles},\ }\href {https://doi.org/10.1038/s42254-020-0224-2} {\bibfield  {journal} {\bibinfo  {journal} {Nat. Rev. Phys.}\ }\textbf {\bibinfo {volume} {2}},\ \bibinfo {pages} {538} (\bibinfo {year} {2020})}\BibitemShut {NoStop}%
\bibitem [{\citenamefont {Degen}\ \emph {et~al.}(2017)\citenamefont {Degen}, \citenamefont {Reinhard},\ and\ \citenamefont {Cappellaro}}]{Degen_2017}%
  \BibitemOpen
  \bibfield  {author} {\bibinfo {author} {\bibfnamefont {C.~L.}\ \bibnamefont {Degen}}, \bibinfo {author} {\bibfnamefont {F.}~\bibnamefont {Reinhard}},\ and\ \bibinfo {author} {\bibfnamefont {P.}~\bibnamefont {Cappellaro}},\ }\bibfield  {title} {\bibinfo {title} {Quantum sensing},\ }\href {https://doi.org/10.1103/RevModPhys.89.035002} {\bibfield  {journal} {\bibinfo  {journal} {Rev. Mod. Phys.}\ }\textbf {\bibinfo {volume} {89}},\ \bibinfo {pages} {035002} (\bibinfo {year} {2017})}\BibitemShut {NoStop}%
\bibitem [{\citenamefont {Karnieli}\ \emph {et~al.}(2023)\citenamefont {Karnieli}, \citenamefont {Tsesses}, \citenamefont {Yu}, \citenamefont {Rivera}, \citenamefont {Zhao}, \citenamefont {Arie}, \citenamefont {Fan},\ and\ \citenamefont {Kaminer}}]{Karnieli_2023}%
  \BibitemOpen
  \bibfield  {author} {\bibinfo {author} {\bibfnamefont {A.}~\bibnamefont {Karnieli}}, \bibinfo {author} {\bibfnamefont {S.}~\bibnamefont {Tsesses}}, \bibinfo {author} {\bibfnamefont {R.}~\bibnamefont {Yu}}, \bibinfo {author} {\bibfnamefont {N.}~\bibnamefont {Rivera}}, \bibinfo {author} {\bibfnamefont {Z.}~\bibnamefont {Zhao}}, \bibinfo {author} {\bibfnamefont {A.}~\bibnamefont {Arie}}, \bibinfo {author} {\bibfnamefont {S.}~\bibnamefont {Fan}},\ and\ \bibinfo {author} {\bibfnamefont {I.}~\bibnamefont {Kaminer}},\ }\bibfield  {title} {\bibinfo {title} {Quantum sensing of strongly coupled light-matter systems using free electrons},\ }\href {https://doi.org/10.1126/sciadv.add2349} {\bibfield  {journal} {\bibinfo  {journal} {Sci. Adv.}\ }\textbf {\bibinfo {volume} {9}},\ \bibinfo {pages} {eadd2349} (\bibinfo {year} {2023})}\BibitemShut {NoStop}%
\bibitem [{\citenamefont {Joo}\ \emph {et~al.}(2011)\citenamefont {Joo}, \citenamefont {Munro},\ and\ \citenamefont {Spiller}}]{Hoo_2011}%
  \BibitemOpen
  \bibfield  {author} {\bibinfo {author} {\bibfnamefont {J.}~\bibnamefont {Joo}}, \bibinfo {author} {\bibfnamefont {W.~J.}\ \bibnamefont {Munro}},\ and\ \bibinfo {author} {\bibfnamefont {T.~P.}\ \bibnamefont {Spiller}},\ }\bibfield  {title} {\bibinfo {title} {Quantum metrology with entangled coherent states},\ }\href {https://doi.org/10.1103/PhysRevLett.107.083601} {\bibfield  {journal} {\bibinfo  {journal} {Phys. Rev. Lett.}\ }\textbf {\bibinfo {volume} {107}},\ \bibinfo {pages} {083601} (\bibinfo {year} {2011})}\BibitemShut {NoStop}%
\bibitem [{\citenamefont {Bai}\ and\ \citenamefont {An}(2023)}]{Bai_2023}%
  \BibitemOpen
  \bibfield  {author} {\bibinfo {author} {\bibfnamefont {S.-Y.}\ \bibnamefont {Bai}}\ and\ \bibinfo {author} {\bibfnamefont {J.-H.}\ \bibnamefont {An}},\ }\bibfield  {title} {\bibinfo {title} {Floquet engineering to overcome no-go theorem of noisy quantum metrology},\ }\href {https://doi.org/10.1103/PhysRevLett.131.050801} {\bibfield  {journal} {\bibinfo  {journal} {Phys. Rev. Lett.}\ }\textbf {\bibinfo {volume} {131}},\ \bibinfo {pages} {050801} (\bibinfo {year} {2023})}\BibitemShut {NoStop}%
\bibitem [{\citenamefont {Pirmoradian}\ and\ \citenamefont {M\o{}lmer}(2019)}]{Faezeh_2019}%
  \BibitemOpen
  \bibfield  {author} {\bibinfo {author} {\bibfnamefont {F.}~\bibnamefont {Pirmoradian}}\ and\ \bibinfo {author} {\bibfnamefont {K.}~\bibnamefont {M\o{}lmer}},\ }\bibfield  {title} {\bibinfo {title} {Aging of a quantum battery},\ }\href {https://doi.org/10.1103/PhysRevA.100.043833} {\bibfield  {journal} {\bibinfo  {journal} {Phys. Rev. A}\ }\textbf {\bibinfo {volume} {100}},\ \bibinfo {pages} {043833} (\bibinfo {year} {2019})}\BibitemShut {NoStop}%
\bibitem [{\citenamefont {Zhang}\ \emph {et~al.}(2019)\citenamefont {Zhang}, \citenamefont {Yang}, \citenamefont {Fu},\ and\ \citenamefont {Wang}}]{Zhangyy_2019}%
  \BibitemOpen
  \bibfield  {author} {\bibinfo {author} {\bibfnamefont {Y.-Y.}\ \bibnamefont {Zhang}}, \bibinfo {author} {\bibfnamefont {T.-R.}\ \bibnamefont {Yang}}, \bibinfo {author} {\bibfnamefont {L.}~\bibnamefont {Fu}},\ and\ \bibinfo {author} {\bibfnamefont {X.}~\bibnamefont {Wang}},\ }\bibfield  {title} {\bibinfo {title} {Powerful harmonic charging in a quantum battery},\ }\href {https://doi.org/10.1103/PhysRevE.99.052106} {\bibfield  {journal} {\bibinfo  {journal} {Phys. Rev. E}\ }\textbf {\bibinfo {volume} {99}},\ \bibinfo {pages} {052106} (\bibinfo {year} {2019})}\BibitemShut {NoStop}%
\bibitem [{\citenamefont {Crescente}\ \emph {et~al.}(2020)\citenamefont {Crescente}, \citenamefont {Carrega}, \citenamefont {Sassetti},\ and\ \citenamefont {Ferraro}}]{Crescente_2020}%
  \BibitemOpen
  \bibfield  {author} {\bibinfo {author} {\bibfnamefont {A.}~\bibnamefont {Crescente}}, \bibinfo {author} {\bibfnamefont {M.}~\bibnamefont {Carrega}}, \bibinfo {author} {\bibfnamefont {M.}~\bibnamefont {Sassetti}},\ and\ \bibinfo {author} {\bibfnamefont {D.}~\bibnamefont {Ferraro}},\ }\bibfield  {title} {\bibinfo {title} {Charging and energy fluctuations of a driven quantum battery},\ }\href {https://doi.org/10.1088/1367-2630/ab91fc} {\bibfield  {journal} {\bibinfo  {journal} {New J. Phys.}\ }\textbf {\bibinfo {volume} {22}},\ \bibinfo {pages} {063057} (\bibinfo {year} {2020})}\BibitemShut {NoStop}%
\bibitem [{\citenamefont {Arjmandi}\ \emph {et~al.}(2022)\citenamefont {Arjmandi}, \citenamefont {Shokri}, \citenamefont {Faizi},\ and\ \citenamefont {Mohammadi}}]{Arjmandi}%
  \BibitemOpen
  \bibfield  {author} {\bibinfo {author} {\bibfnamefont {M.~B.}\ \bibnamefont {Arjmandi}}, \bibinfo {author} {\bibfnamefont {A.}~\bibnamefont {Shokri}}, \bibinfo {author} {\bibfnamefont {E.}~\bibnamefont {Faizi}},\ and\ \bibinfo {author} {\bibfnamefont {H.}~\bibnamefont {Mohammadi}},\ }\bibfield  {title} {\bibinfo {title} {Performance of quantum batteries with correlated and uncorrelated chargers},\ }\href {https://doi.org/10.1103/PhysRevA.106.062609} {\bibfield  {journal} {\bibinfo  {journal} {Phys. Rev. A}\ }\textbf {\bibinfo {volume} {106}},\ \bibinfo {pages} {062609} (\bibinfo {year} {2022})}\BibitemShut {NoStop}%
\bibitem [{\citenamefont {Braak}(2011)}]{Braak_2011}%
  \BibitemOpen
  \bibfield  {author} {\bibinfo {author} {\bibfnamefont {D.}~\bibnamefont {Braak}},\ }\bibfield  {title} {\bibinfo {title} {Integrability of the rabi model},\ }\href {https://doi.org/10.1103/PhysRevLett.107.100401} {\bibfield  {journal} {\bibinfo  {journal} {Phys. Rev. Lett.}\ }\textbf {\bibinfo {volume} {107}},\ \bibinfo {pages} {100401} (\bibinfo {year} {2011})}\BibitemShut {NoStop}%
\bibitem [{\citenamefont {Braak}\ \emph {et~al.}(2016)\citenamefont {Braak}, \citenamefont {Chen}, \citenamefont {Batchelor},\ and\ \citenamefont {Solano}}]{Braak_2016}%
  \BibitemOpen
  \bibfield  {author} {\bibinfo {author} {\bibfnamefont {D.}~\bibnamefont {Braak}}, \bibinfo {author} {\bibfnamefont {Q.-H.}\ \bibnamefont {Chen}}, \bibinfo {author} {\bibfnamefont {M.~T.}\ \bibnamefont {Batchelor}},\ and\ \bibinfo {author} {\bibfnamefont {E.}~\bibnamefont {Solano}},\ }\bibfield  {title} {\bibinfo {title} {Semi-classical and quantum rabi models: in celebration of 80 years},\ }\href {https://doi.org/10.1088/1751-8113/49/30/300301} {\bibfield  {journal} {\bibinfo  {journal} {J. Phys. A: Math. Theor.}\ }\textbf {\bibinfo {volume} {49}},\ \bibinfo {pages} {300301} (\bibinfo {year} {2016})}\BibitemShut {NoStop}%
\bibitem [{\citenamefont {Xie}\ \emph {et~al.}(2017)\citenamefont {Xie}, \citenamefont {Zhong}, \citenamefont {Batchelor},\ and\ \citenamefont {Lee}}]{Xie_2017}%
  \BibitemOpen
  \bibfield  {author} {\bibinfo {author} {\bibfnamefont {Q.}~\bibnamefont {Xie}}, \bibinfo {author} {\bibfnamefont {H.}~\bibnamefont {Zhong}}, \bibinfo {author} {\bibfnamefont {M.~T.}\ \bibnamefont {Batchelor}},\ and\ \bibinfo {author} {\bibfnamefont {C.}~\bibnamefont {Lee}},\ }\bibfield  {title} {\bibinfo {title} {The quantum rabi model: solution and dynamics},\ }\href {https://doi.org/10.1088/1751-8121/aa5a65} {\bibfield  {journal} {\bibinfo  {journal} {J. Phys. A: Math. Theor.}\ }\textbf {\bibinfo {volume} {50}},\ \bibinfo {pages} {113001} (\bibinfo {year} {2017})}\BibitemShut {NoStop}%
\bibitem [{\citenamefont {Yoshihara}\ \emph {et~al.}(2017)\citenamefont {Yoshihara}, \citenamefont {Fuse}, \citenamefont {Ashhab}, \citenamefont {Kakuyanagi}, \citenamefont {Saito},\ and\ \citenamefont {Semba}}]{Yoshihara_2016}%
  \BibitemOpen
  \bibfield  {author} {\bibinfo {author} {\bibfnamefont {F.}~\bibnamefont {Yoshihara}}, \bibinfo {author} {\bibfnamefont {T.}~\bibnamefont {Fuse}}, \bibinfo {author} {\bibfnamefont {S.}~\bibnamefont {Ashhab}}, \bibinfo {author} {\bibfnamefont {K.}~\bibnamefont {Kakuyanagi}}, \bibinfo {author} {\bibfnamefont {S.}~\bibnamefont {Saito}},\ and\ \bibinfo {author} {\bibfnamefont {K.}~\bibnamefont {Semba}},\ }\bibfield  {title} {\bibinfo {title} {Superconducting qubit-oscillator circuit beyond the ultrastrong-coupling regime},\ }\href {https://doi.org/10.1038/nphys3906} {\bibfield  {journal} {\bibinfo  {journal} {Nat. Phys.}\ }\textbf {\bibinfo {volume} {13}},\ \bibinfo {pages} {44} (\bibinfo {year} {2017})}\BibitemShut {NoStop}%
\bibitem [{\citenamefont {Felicetti}\ \emph {et~al.}(2017)\citenamefont {Felicetti}, \citenamefont {Rico}, \citenamefont {Sabin}, \citenamefont {Ockenfels}, \citenamefont {Koch}, \citenamefont {Leder}, \citenamefont {Grossert}, \citenamefont {Weitz},\ and\ \citenamefont {Solano}}]{Simone_2017}%
  \BibitemOpen
  \bibfield  {author} {\bibinfo {author} {\bibfnamefont {S.}~\bibnamefont {Felicetti}}, \bibinfo {author} {\bibfnamefont {E.}~\bibnamefont {Rico}}, \bibinfo {author} {\bibfnamefont {C.}~\bibnamefont {Sabin}}, \bibinfo {author} {\bibfnamefont {T.}~\bibnamefont {Ockenfels}}, \bibinfo {author} {\bibfnamefont {J.}~\bibnamefont {Koch}}, \bibinfo {author} {\bibfnamefont {M.}~\bibnamefont {Leder}}, \bibinfo {author} {\bibfnamefont {C.}~\bibnamefont {Grossert}}, \bibinfo {author} {\bibfnamefont {M.}~\bibnamefont {Weitz}},\ and\ \bibinfo {author} {\bibfnamefont {E.}~\bibnamefont {Solano}},\ }\bibfield  {title} {\bibinfo {title} {Quantum rabi model in the brillouin zone with ultracold atoms},\ }\href {https://doi.org/10.1103/PhysRevA.95.013827} {\bibfield  {journal} {\bibinfo  {journal} {Phys. Rev. A}\ }\textbf {\bibinfo {volume} {95}},\ \bibinfo {pages} {013827} (\bibinfo {year} {2017})}\BibitemShut {NoStop}%
\bibitem [{\citenamefont {Lv}\ \emph {et~al.}(2018)\citenamefont {Lv}, \citenamefont {An}, \citenamefont {Liu}, \citenamefont {Zhang}, \citenamefont {Pedernales}, \citenamefont {Lamata}, \citenamefont {Solano},\ and\ \citenamefont {Kim}}]{Lv_2018}%
  \BibitemOpen
  \bibfield  {author} {\bibinfo {author} {\bibfnamefont {D.}~\bibnamefont {Lv}}, \bibinfo {author} {\bibfnamefont {S.}~\bibnamefont {An}}, \bibinfo {author} {\bibfnamefont {Z.}~\bibnamefont {Liu}}, \bibinfo {author} {\bibfnamefont {J.-N.}\ \bibnamefont {Zhang}}, \bibinfo {author} {\bibfnamefont {J.~S.}\ \bibnamefont {Pedernales}}, \bibinfo {author} {\bibfnamefont {L.}~\bibnamefont {Lamata}}, \bibinfo {author} {\bibfnamefont {E.}~\bibnamefont {Solano}},\ and\ \bibinfo {author} {\bibfnamefont {K.}~\bibnamefont {Kim}},\ }\bibfield  {title} {\bibinfo {title} {Quantum simulation of the quantum rabi model in a trapped ion},\ }\href {https://doi.org/10.1103/PhysRevX.8.021027} {\bibfield  {journal} {\bibinfo  {journal} {Phys. Rev. X}\ }\textbf {\bibinfo {volume} {8}},\ \bibinfo {pages} {021027} (\bibinfo {year} {2018})}\BibitemShut {NoStop}%
\bibitem [{\citenamefont {Kockum}\ \emph {et~al.}(2019)\citenamefont {Kockum}, \citenamefont {Miranowicz}, \citenamefont {Liberato}, \citenamefont {Savasta},\ and\ \citenamefont {Nori}}]{Frisk_Kockum_2019}%
  \BibitemOpen
  \bibfield  {author} {\bibinfo {author} {\bibfnamefont {A.~F.}\ \bibnamefont {Kockum}}, \bibinfo {author} {\bibfnamefont {A.}~\bibnamefont {Miranowicz}}, \bibinfo {author} {\bibfnamefont {S.~D.}\ \bibnamefont {Liberato}}, \bibinfo {author} {\bibfnamefont {S.}~\bibnamefont {Savasta}},\ and\ \bibinfo {author} {\bibfnamefont {F.}~\bibnamefont {Nori}},\ }\bibfield  {title} {\bibinfo {title} {Ultrastrong coupling between light and matter},\ }\href {https://doi.org/https://doi.org/10.1038/s42254-018-0006-2} {\bibfield  {journal} {\bibinfo  {journal} {Nat. Rev. Phys.}\ }\textbf {\bibinfo {volume} {1}},\ \bibinfo {pages} {19} (\bibinfo {year} {2019})}\BibitemShut {NoStop}%
\bibitem [{\citenamefont {Forn-D{\'{\i}}az}\ \emph {et~al.}(2019)\citenamefont {Forn-D{\'{\i}}az}, \citenamefont {Lamata}, \citenamefont {Rico}, \citenamefont {Kono},\ and\ \citenamefont {Solano}}]{Forn_D_az_2019}%
  \BibitemOpen
  \bibfield  {author} {\bibinfo {author} {\bibfnamefont {P.}~\bibnamefont {Forn-D{\'{\i}}az}}, \bibinfo {author} {\bibfnamefont {L.}~\bibnamefont {Lamata}}, \bibinfo {author} {\bibfnamefont {E.}~\bibnamefont {Rico}}, \bibinfo {author} {\bibfnamefont {J.}~\bibnamefont {Kono}},\ and\ \bibinfo {author} {\bibfnamefont {E.}~\bibnamefont {Solano}},\ }\bibfield  {title} {\bibinfo {title} {Ultrastrong coupling regimes of light-matter interaction},\ }\href {https://doi.org/https://doi.org/10.1103/RevModPhys.91.025005} {\bibfield  {journal} {\bibinfo  {journal} {Rev. Mod. Phys.}\ }\textbf {\bibinfo {volume} {91}},\ \bibinfo {pages} {025005} (\bibinfo {year} {2019})}\BibitemShut {NoStop}%
\bibitem [{\citenamefont {Le~Boit\'{e}}(2020)}]{Le_Boit_2020}%
  \BibitemOpen
  \bibfield  {author} {\bibinfo {author} {\bibfnamefont {A.}~\bibnamefont {Le~Boit\'{e}}},\ }\bibfield  {title} {\bibinfo {title} {Theoretical methods for ultrastrong light-matter interactions},\ }\href {https://doi.org/10.1002/qute.201900140} {\bibfield  {journal} {\bibinfo  {journal} {Adv. Quantum Technol.}\ }\textbf {\bibinfo {volume} {3}},\ \bibinfo {pages} {1900140} (\bibinfo {year} {2020})}\BibitemShut {NoStop}%
\bibitem [{\citenamefont {Yoshihara}\ \emph {et~al.}(2014)\citenamefont {Yoshihara}, \citenamefont {Nakamura}, \citenamefont {Yan}, \citenamefont {Gustavsson}, \citenamefont {Bylander}, \citenamefont {Oliver},\ and\ \citenamefont {Tsai}}]{Yoshihara_2014}%
  \BibitemOpen
  \bibfield  {author} {\bibinfo {author} {\bibfnamefont {F.}~\bibnamefont {Yoshihara}}, \bibinfo {author} {\bibfnamefont {Y.}~\bibnamefont {Nakamura}}, \bibinfo {author} {\bibfnamefont {F.}~\bibnamefont {Yan}}, \bibinfo {author} {\bibfnamefont {S.}~\bibnamefont {Gustavsson}}, \bibinfo {author} {\bibfnamefont {J.}~\bibnamefont {Bylander}}, \bibinfo {author} {\bibfnamefont {W.~D.}\ \bibnamefont {Oliver}},\ and\ \bibinfo {author} {\bibfnamefont {J.-S.}\ \bibnamefont {Tsai}},\ }\bibfield  {title} {\bibinfo {title} {Flux qubit noise spectroscopy using rabi oscillations under strong driving conditions},\ }\href {https://doi.org/10.1103/PhysRevB.89.020503} {\bibfield  {journal} {\bibinfo  {journal} {Phys. Rev. B}\ }\textbf {\bibinfo {volume} {89}},\ \bibinfo {pages} {020503} (\bibinfo {year} {2014})}\BibitemShut {NoStop}%
\bibitem [{\citenamefont {Deng}\ \emph {et~al.}(2015)\citenamefont {Deng}, \citenamefont {Orgiazzi}, \citenamefont {Shen}, \citenamefont {Ashhab},\ and\ \citenamefont {Lupascu}}]{Deng_2015}%
  \BibitemOpen
  \bibfield  {author} {\bibinfo {author} {\bibfnamefont {C.}~\bibnamefont {Deng}}, \bibinfo {author} {\bibfnamefont {J.-L.}\ \bibnamefont {Orgiazzi}}, \bibinfo {author} {\bibfnamefont {F.}~\bibnamefont {Shen}}, \bibinfo {author} {\bibfnamefont {S.}~\bibnamefont {Ashhab}},\ and\ \bibinfo {author} {\bibfnamefont {A.}~\bibnamefont {Lupascu}},\ }\bibfield  {title} {\bibinfo {title} {Observation of floquet states in a strongly driven artificial atom},\ }\href {https://doi.org/10.1103/physrevlett.115.133601} {\bibfield  {journal} {\bibinfo  {journal} {Phys. Rev. Lett.}\ }\textbf {\bibinfo {volume} {115}},\ \bibinfo {pages} {133601} (\bibinfo {year} {2015})}\BibitemShut {NoStop}%
\bibitem [{\citenamefont {Romero}\ \emph {et~al.}(2012)\citenamefont {Romero}, \citenamefont {Ballester}, \citenamefont {Wang}, \citenamefont {Scarani},\ and\ \citenamefont {Solano}}]{Romero_2012}%
  \BibitemOpen
  \bibfield  {author} {\bibinfo {author} {\bibfnamefont {G.}~\bibnamefont {Romero}}, \bibinfo {author} {\bibfnamefont {D.}~\bibnamefont {Ballester}}, \bibinfo {author} {\bibfnamefont {Y.~M.}\ \bibnamefont {Wang}}, \bibinfo {author} {\bibfnamefont {V.}~\bibnamefont {Scarani}},\ and\ \bibinfo {author} {\bibfnamefont {E.}~\bibnamefont {Solano}},\ }\bibfield  {title} {\bibinfo {title} {Ultrafast quantum gates in circuit qed},\ }\href {https://doi.org/10.1103/PhysRevLett.108.120501} {\bibfield  {journal} {\bibinfo  {journal} {Phys. Rev. Lett.}\ }\textbf {\bibinfo {volume} {108}},\ \bibinfo {pages} {120501} (\bibinfo {year} {2012})}\BibitemShut {NoStop}%
\bibitem [{\citenamefont {Song}\ \emph {et~al.}(2016)\citenamefont {Song}, \citenamefont {Kestner}, \citenamefont {Wang},\ and\ \citenamefont {Das~Sarma}}]{Song_2016}%
  \BibitemOpen
  \bibfield  {author} {\bibinfo {author} {\bibfnamefont {Y.}~\bibnamefont {Song}}, \bibinfo {author} {\bibfnamefont {J.~P.}\ \bibnamefont {Kestner}}, \bibinfo {author} {\bibfnamefont {X.}~\bibnamefont {Wang}},\ and\ \bibinfo {author} {\bibfnamefont {S.}~\bibnamefont {Das~Sarma}},\ }\bibfield  {title} {\bibinfo {title} {Fast control of semiconductor qubits beyond the rotating-wave approximation},\ }\href {https://doi.org/10.1103/PhysRevA.94.012321} {\bibfield  {journal} {\bibinfo  {journal} {Phys. Rev. A}\ }\textbf {\bibinfo {volume} {94}},\ \bibinfo {pages} {012321} (\bibinfo {year} {2016})}\BibitemShut {NoStop}%
\bibitem [{\citenamefont {Twyeffort~Irish}\ and\ \citenamefont {Armour}(2022)}]{irish_2022}%
  \BibitemOpen
  \bibfield  {author} {\bibinfo {author} {\bibfnamefont {E.~K.}\ \bibnamefont {Twyeffort~Irish}}\ and\ \bibinfo {author} {\bibfnamefont {A.~D.}\ \bibnamefont {Armour}},\ }\bibfield  {title} {\bibinfo {title} {Defining the semiclassical limit of the quantum rabi hamiltonian},\ }\href {https://doi.org/10.1103/PhysRevLett.129.183603} {\bibfield  {journal} {\bibinfo  {journal} {Phys. Rev. Lett.}\ }\textbf {\bibinfo {volume} {129}},\ \bibinfo {pages} {183603} (\bibinfo {year} {2022})}\BibitemShut {NoStop}%
\bibitem [{\citenamefont {Hausinger}\ and\ \citenamefont {Grifoni}(2010)}]{hausinger_2010}%
  \BibitemOpen
  \bibfield  {author} {\bibinfo {author} {\bibfnamefont {J.}~\bibnamefont {Hausinger}}\ and\ \bibinfo {author} {\bibfnamefont {M.}~\bibnamefont {Grifoni}},\ }\bibfield  {title} {\bibinfo {title} {Qubit-oscillator system: An analytical treatment of the ultrastrong coupling regime},\ }\href {https://doi.org/10.1103/PhysRevA.82.062320} {\bibfield  {journal} {\bibinfo  {journal} {Phys. Rev. A}\ }\textbf {\bibinfo {volume} {82}},\ \bibinfo {pages} {062320} (\bibinfo {year} {2010})}\BibitemShut {NoStop}%
\bibitem [{\citenamefont {Gan}\ and\ \citenamefont {Zheng}(2010)}]{Gan_2010}%
  \BibitemOpen
  \bibfield  {author} {\bibinfo {author} {\bibfnamefont {C.~J.}\ \bibnamefont {Gan}}\ and\ \bibinfo {author} {\bibfnamefont {H.}~\bibnamefont {Zheng}},\ }\bibfield  {title} {\bibinfo {title} {Dynamics of a two-level system coupled to a quantum oscillator: transformed rotating-wave approximation},\ }\href {https://doi.org/10.1140/epjd/e2010-00182-8} {\bibfield  {journal} {\bibinfo  {journal} {Eur. Phys. J. D}\ }\textbf {\bibinfo {volume} {59}},\ \bibinfo {pages} {473} (\bibinfo {year} {2010})}\BibitemShut {NoStop}%
\bibitem [{\citenamefont {Zhang}(2016)}]{Zhang_2016}%
  \BibitemOpen
  \bibfield  {author} {\bibinfo {author} {\bibfnamefont {Y.-Y.}\ \bibnamefont {Zhang}},\ }\bibfield  {title} {\bibinfo {title} {Generalized squeezing rotating-wave approximation to the isotropic and anisotropic rabi model in the ultrastrong-coupling regime},\ }\href {https://doi.org/10.1103/PhysRevA.94.063824} {\bibfield  {journal} {\bibinfo  {journal} {Phys. Rev. A}\ }\textbf {\bibinfo {volume} {94}},\ \bibinfo {pages} {063824} (\bibinfo {year} {2016})}\BibitemShut {NoStop}%
\bibitem [{\citenamefont {Cong}\ \emph {et~al.}(2019)\citenamefont {Cong}, \citenamefont {Sun}, \citenamefont {Liu}, \citenamefont {Ying},\ and\ \citenamefont {Luo}}]{Lei_2019}%
  \BibitemOpen
  \bibfield  {author} {\bibinfo {author} {\bibfnamefont {L.}~\bibnamefont {Cong}}, \bibinfo {author} {\bibfnamefont {X.-M.}\ \bibnamefont {Sun}}, \bibinfo {author} {\bibfnamefont {M.}~\bibnamefont {Liu}}, \bibinfo {author} {\bibfnamefont {Z.-J.}\ \bibnamefont {Ying}},\ and\ \bibinfo {author} {\bibfnamefont {H.-G.}\ \bibnamefont {Luo}},\ }\bibfield  {title} {\bibinfo {title} {Polaron picture of the two-photon quantum rabi model},\ }\href {https://doi.org/10.1103/PhysRevA.99.013815} {\bibfield  {journal} {\bibinfo  {journal} {Phys. Rev. A}\ }\textbf {\bibinfo {volume} {99}},\ \bibinfo {pages} {013815} (\bibinfo {year} {2019})}\BibitemShut {NoStop}%
\bibitem [{\citenamefont {Werther}\ and\ \citenamefont {Grossmann}(2017)}]{Werther_2017}%
  \BibitemOpen
  \bibfield  {author} {\bibinfo {author} {\bibfnamefont {M.}~\bibnamefont {Werther}}\ and\ \bibinfo {author} {\bibfnamefont {F.}~\bibnamefont {Grossmann}},\ }\bibfield  {title} {\bibinfo {title} {Including temperature in a wavefunction description of the dynamics of the quantum rabi model},\ }\href {https://doi.org/10.1088/1751-8121/aa94a1} {\bibfield  {journal} {\bibinfo  {journal} {J. Phys. A: Math. Theor.}\ }\textbf {\bibinfo {volume} {51}},\ \bibinfo {pages} {014001} (\bibinfo {year} {2017})}\BibitemShut {NoStop}%
\bibitem [{\citenamefont {Werther}\ and\ \citenamefont {Grossmann}(2018)}]{Werther_2018}%
  \BibitemOpen
  \bibfield  {author} {\bibinfo {author} {\bibfnamefont {M.}~\bibnamefont {Werther}}\ and\ \bibinfo {author} {\bibfnamefont {F.}~\bibnamefont {Grossmann}},\ }\bibfield  {title} {\bibinfo {title} {The davydov d1.5 ansatz for the quantum rabi model},\ }\href {https://doi.org/10.1088/1402-4896/aac7f9} {\bibfield  {journal} {\bibinfo  {journal} {Physica Scripta}\ }\textbf {\bibinfo {volume} {93}},\ \bibinfo {pages} {074001} (\bibinfo {year} {2018})}\BibitemShut {NoStop}%
\bibitem [{\citenamefont {Li}\ \emph {et~al.}(2021)\citenamefont {Li}, \citenamefont {Ferri},\ and\ \citenamefont {Batchelor}}]{Lizm_2021}%
  \BibitemOpen
  \bibfield  {author} {\bibinfo {author} {\bibfnamefont {Z.-M.}\ \bibnamefont {Li}}, \bibinfo {author} {\bibfnamefont {D.}~\bibnamefont {Ferri}},\ and\ \bibinfo {author} {\bibfnamefont {M.~T.}\ \bibnamefont {Batchelor}},\ }\bibfield  {title} {\bibinfo {title} {Nonorthogonal-qubit-state expansion for the asymmetric quantum rabi model},\ }\href {https://doi.org/10.1103/PhysRevA.103.013711} {\bibfield  {journal} {\bibinfo  {journal} {Phys. Rev. A}\ }\textbf {\bibinfo {volume} {103}},\ \bibinfo {pages} {013711} (\bibinfo {year} {2021})}\BibitemShut {NoStop}%
\bibitem [{\citenamefont {Chen}\ \emph {et~al.}(2012)\citenamefont {Chen}, \citenamefont {Wang}, \citenamefont {He}, \citenamefont {Liu},\ and\ \citenamefont {Wang}}]{chen_2012}%
  \BibitemOpen
  \bibfield  {author} {\bibinfo {author} {\bibfnamefont {Q.-H.}\ \bibnamefont {Chen}}, \bibinfo {author} {\bibfnamefont {C.}~\bibnamefont {Wang}}, \bibinfo {author} {\bibfnamefont {S.}~\bibnamefont {He}}, \bibinfo {author} {\bibfnamefont {T.}~\bibnamefont {Liu}},\ and\ \bibinfo {author} {\bibfnamefont {K.-L.}\ \bibnamefont {Wang}},\ }\bibfield  {title} {\bibinfo {title} {Exact solvability of the quantum rabi model using bogoliubov operators},\ }\href {https://doi.org/10.1103/PhysRevA.86.023822} {\bibfield  {journal} {\bibinfo  {journal} {Phys. Rev. A}\ }\textbf {\bibinfo {volume} {86}},\ \bibinfo {pages} {023822} (\bibinfo {year} {2012})}\BibitemShut {NoStop}%
\bibitem [{\citenamefont {Shirley}(1965)}]{Shirley_1965}%
  \BibitemOpen
  \bibfield  {author} {\bibinfo {author} {\bibfnamefont {J.~H.}\ \bibnamefont {Shirley}},\ }\bibfield  {title} {\bibinfo {title} {Solution of the schr\"{o}dinger equation with a hamiltonian periodic in time},\ }\href {https://doi.org/https://doi.org/10.1103/PhysRev.138.B979} {\bibfield  {journal} {\bibinfo  {journal} {Phys. Rev.}\ }\textbf {\bibinfo {volume} {138}},\ \bibinfo {pages} {B979} (\bibinfo {year} {1965})}\BibitemShut {NoStop}%
\bibitem [{\citenamefont {Chu}\ and\ \citenamefont {Telnov}(2004)}]{Chu_2004}%
  \BibitemOpen
  \bibfield  {author} {\bibinfo {author} {\bibfnamefont {S.-I.}\ \bibnamefont {Chu}}\ and\ \bibinfo {author} {\bibfnamefont {D.~A.}\ \bibnamefont {Telnov}},\ }\bibfield  {title} {\bibinfo {title} {Beyond the floquet theorem: generalized floquet formalisms and quasienergy methods for atomic and molecular multiphoton processes in intense laser fields},\ }\href {https://doi.org/https://doi.org/10.1016/j.physrep.2003.10.001} {\bibfield  {journal} {\bibinfo  {journal} {Phys. Rep.}\ }\textbf {\bibinfo {volume} {390}},\ \bibinfo {pages} {1} (\bibinfo {year} {2004})}\BibitemShut {NoStop}%
\bibitem [{\citenamefont {Engelhardt}\ \emph {et~al.}(2024)\citenamefont {Engelhardt}, \citenamefont {Choudhury},\ and\ \citenamefont {Liu}}]{Georg_2024}%
  \BibitemOpen
  \bibfield  {author} {\bibinfo {author} {\bibfnamefont {G.}~\bibnamefont {Engelhardt}}, \bibinfo {author} {\bibfnamefont {S.}~\bibnamefont {Choudhury}},\ and\ \bibinfo {author} {\bibfnamefont {W.~V.}\ \bibnamefont {Liu}},\ }\bibfield  {title} {\bibinfo {title} {Unified light-matter floquet theory and its application to quantum communication},\ }\href {https://doi.org/10.1103/PhysRevResearch.6.013116} {\bibfield  {journal} {\bibinfo  {journal} {Phys. Rev. Res.}\ }\textbf {\bibinfo {volume} {6}},\ \bibinfo {pages} {013116} (\bibinfo {year} {2024})}\BibitemShut {NoStop}%
\bibitem [{\citenamefont {Zhao}\ \emph {et~al.}(2022)\citenamefont {Zhao}, \citenamefont {Sun}, \citenamefont {Chen},\ and\ \citenamefont {Gelin}}]{Zhao_2022}%
  \BibitemOpen
  \bibfield  {author} {\bibinfo {author} {\bibfnamefont {Y.}~\bibnamefont {Zhao}}, \bibinfo {author} {\bibfnamefont {K.}~\bibnamefont {Sun}}, \bibinfo {author} {\bibfnamefont {L.}~\bibnamefont {Chen}},\ and\ \bibinfo {author} {\bibfnamefont {M.}~\bibnamefont {Gelin}},\ }\bibfield  {title} {\bibinfo {title} {The hierarchy of davydovs ans\"{a}tze and its applications},\ }\href {https://doi.org/10.1002/wcms.1589} {\bibfield  {journal} {\bibinfo  {journal} {{WIREs} Comput. Mol. Sci.}\ }\textbf {\bibinfo {volume} {12}},\ \bibinfo {pages} {e1589} (\bibinfo {year} {2022})}\BibitemShut {NoStop}%
\bibitem [{\citenamefont {Zhao}(2023)}]{Zhao_2023}%
  \BibitemOpen
  \bibfield  {author} {\bibinfo {author} {\bibfnamefont {Y.}~\bibnamefont {Zhao}},\ }\bibfield  {title} {\bibinfo {title} {The hierarchy of davydov's ans\"{a}tze: From guesswork to numerically {\textquotedblleft}exact{\textquotedblright} many-body wave functions},\ }\href {https://doi.org/10.1063/5.0140002} {\bibfield  {journal} {\bibinfo  {journal} {J. Chem. Phys.}\ }\textbf {\bibinfo {volume} {158}},\ \bibinfo {pages} {080901} (\bibinfo {year} {2023})}\BibitemShut {NoStop}%
\bibitem [{\citenamefont {Wang}\ \emph {et~al.}(2017)\citenamefont {Wang}, \citenamefont {Fujihashi}, \citenamefont {Chen},\ and\ \citenamefont {Zhao}}]{WangLu_2017}%
  \BibitemOpen
  \bibfield  {author} {\bibinfo {author} {\bibfnamefont {L.}~\bibnamefont {Wang}}, \bibinfo {author} {\bibfnamefont {Y.}~\bibnamefont {Fujihashi}}, \bibinfo {author} {\bibfnamefont {L.}~\bibnamefont {Chen}},\ and\ \bibinfo {author} {\bibfnamefont {Y.}~\bibnamefont {Zhao}},\ }\bibfield  {title} {\bibinfo {title} {Finite-temperature time-dependent variation with multiple davydov states},\ }\href {https://doi.org/10.1063/1.4979017} {\bibfield  {journal} {\bibinfo  {journal} {J. Chem. Phys.}\ }\textbf {\bibinfo {volume} {146}},\ \bibinfo {pages} {124127} (\bibinfo {year} {2017})}\BibitemShut {NoStop}%
\bibitem [{\citenamefont {Werther}\ and\ \citenamefont {Gro{\ss}mann}(2020)}]{Werther_2020}%
  \BibitemOpen
  \bibfield  {author} {\bibinfo {author} {\bibfnamefont {M.}~\bibnamefont {Werther}}\ and\ \bibinfo {author} {\bibfnamefont {F.}~\bibnamefont {Gro{\ss}mann}},\ }\bibfield  {title} {\bibinfo {title} {Apoptosis of moving nonorthogonal basis functions in many-particle quantum dynamics},\ }\href {https://doi.org/10.1103/physrevb.101.174315} {\bibfield  {journal} {\bibinfo  {journal} {Phys. Rev. B}\ }\textbf {\bibinfo {volume} {101}},\ \bibinfo {pages} {174315} (\bibinfo {year} {2020})}\BibitemShut {NoStop}%
\bibitem [{\citenamefont {Yan}(2023)}]{YanPRA_2023}%
  \BibitemOpen
  \bibfield  {author} {\bibinfo {author} {\bibfnamefont {Y.}~\bibnamefont {Yan}},\ }\bibfield  {title} {\bibinfo {title} {Spontaneous emission spectrum from a v-type artificial atom in a strong-coupling regime: Dark lines and line narrowing},\ }\href {https://doi.org/10.1103/PhysRevA.108.043712} {\bibfield  {journal} {\bibinfo  {journal} {Phys. Rev. A}\ }\textbf {\bibinfo {volume} {108}},\ \bibinfo {pages} {043712} (\bibinfo {year} {2023})}\BibitemShut {NoStop}%
\bibitem [{\citenamefont {Frenkel}(1934)}]{frenkel}%
  \BibitemOpen
  \bibfield  {author} {\bibinfo {author} {\bibfnamefont {J.}~\bibnamefont {Frenkel}},\ }\href@noop {} {\emph {\bibinfo {title} {Wave Mechanics}}}\ (\bibinfo  {publisher} {Oxford University Press},\ \bibinfo {address} {Oxford},\ \bibinfo {year} {1934})\BibitemShut {NoStop}%
\bibitem [{\citenamefont {Everitt}\ \emph {et~al.}(2009)\citenamefont {Everitt}, \citenamefont {Munro},\ and\ \citenamefont {Spiller}}]{Everitt_2009}%
  \BibitemOpen
  \bibfield  {author} {\bibinfo {author} {\bibfnamefont {M.~J.}\ \bibnamefont {Everitt}}, \bibinfo {author} {\bibfnamefont {W.~J.}\ \bibnamefont {Munro}},\ and\ \bibinfo {author} {\bibfnamefont {T.~P.}\ \bibnamefont {Spiller}},\ }\bibfield  {title} {\bibinfo {title} {Quantum-classical crossover of a field mode},\ }\href {https://doi.org/10.1103/PhysRevA.79.032328} {\bibfield  {journal} {\bibinfo  {journal} {Phys. Rev. A}\ }\textbf {\bibinfo {volume} {79}},\ \bibinfo {pages} {032328} (\bibinfo {year} {2009})}\BibitemShut {NoStop}%
\bibitem [{\citenamefont {Stenholm}(1972)}]{Stenholm_1972}%
  \BibitemOpen
  \bibfield  {author} {\bibinfo {author} {\bibfnamefont {S.}~\bibnamefont {Stenholm}},\ }\bibfield  {title} {\bibinfo {title} {Saturation effects in rf spectroscopy. i. general theory},\ }\href {https://doi.org/10.1088/0022-3700/5/4/023} {\bibfield  {journal} {\bibinfo  {journal} {J. Phys. B: At. Mol. Phys.}\ }\textbf {\bibinfo {volume} {5}},\ \bibinfo {pages} {878} (\bibinfo {year} {1972})}\BibitemShut {NoStop}%
\bibitem [{\citenamefont {L\"u}\ and\ \citenamefont {Zheng}(2012)}]{Zhiguo_2012}%
  \BibitemOpen
  \bibfield  {author} {\bibinfo {author} {\bibfnamefont {Z.}~\bibnamefont {L\"u}}\ and\ \bibinfo {author} {\bibfnamefont {H.}~\bibnamefont {Zheng}},\ }\bibfield  {title} {\bibinfo {title} {Effects of counter-rotating interaction on driven tunneling dynamics: Coherent destruction of tunneling and bloch-siegert shift},\ }\href {https://doi.org/10.1103/PhysRevA.86.023831} {\bibfield  {journal} {\bibinfo  {journal} {Phys. Rev. A}\ }\textbf {\bibinfo {volume} {86}},\ \bibinfo {pages} {023831} (\bibinfo {year} {2012})}\BibitemShut {NoStop}%
\bibitem [{\citenamefont {Guccione-Gush}\ and\ \citenamefont {Gush}(1974)}]{Guccione_Gush_1974}%
  \BibitemOpen
  \bibfield  {author} {\bibinfo {author} {\bibfnamefont {R.}~\bibnamefont {Guccione-Gush}}\ and\ \bibinfo {author} {\bibfnamefont {H.~P.}\ \bibnamefont {Gush}},\ }\bibfield  {title} {\bibinfo {title} {Two-level system in a bichromatic field},\ }\href {https://doi.org/https://doi.org/10.1103/PhysRevA.10.1474} {\bibfield  {journal} {\bibinfo  {journal} {Phys. Rev. A}\ }\textbf {\bibinfo {volume} {10}},\ \bibinfo {pages} {1474} (\bibinfo {year} {1974})}\BibitemShut {NoStop}%
\bibitem [{\citenamefont {Ho}\ and\ \citenamefont {Chu}(1984)}]{Ho_1984}%
  \BibitemOpen
  \bibfield  {author} {\bibinfo {author} {\bibfnamefont {T.-S.}\ \bibnamefont {Ho}}\ and\ \bibinfo {author} {\bibfnamefont {S.-I.}\ \bibnamefont {Chu}},\ }\bibfield  {title} {\bibinfo {title} {Semiclassical many-mode floquet theory. {II}. non-linear multiphoton dynamics of a two-level system in a strong bichromatic field},\ }\href {https://doi.org/10.1088/0022-3700/17/10/015} {\bibfield  {journal} {\bibinfo  {journal} {J. Phys. B}\ }\textbf {\bibinfo {volume} {17}},\ \bibinfo {pages} {2101} (\bibinfo {year} {1984})}\BibitemShut {NoStop}%
\bibitem [{\citenamefont {Ruyten}(1989)}]{Ruyten_1989}%
  \BibitemOpen
  \bibfield  {author} {\bibinfo {author} {\bibfnamefont {W.~M.}\ \bibnamefont {Ruyten}},\ }\bibfield  {title} {\bibinfo {title} {Harmonic behavior of the multiple quantum resonances of a two-level atom driven by a fully-amplitude-modulated field},\ }\href {https://doi.org/https://doi.org/10.1103/PhysRevA.40.1447} {\bibfield  {journal} {\bibinfo  {journal} {Phys. Rev. A}\ }\textbf {\bibinfo {volume} {40}},\ \bibinfo {pages} {1447} (\bibinfo {year} {1989})}\BibitemShut {NoStop}%
\bibitem [{\citenamefont {Ruyten}(1992)}]{Ruyten_1992}%
  \BibitemOpen
  \bibfield  {author} {\bibinfo {author} {\bibfnamefont {W.~M.}\ \bibnamefont {Ruyten}},\ }\bibfield  {title} {\bibinfo {title} {Resonance behavior of a two-level quantum system in a two-frequency field},\ }\href {https://doi.org/https://doi.org/10.1103/PhysRevA.46.4077} {\bibfield  {journal} {\bibinfo  {journal} {Phys. Rev. A}\ }\textbf {\bibinfo {volume} {46}},\ \bibinfo {pages} {4077} (\bibinfo {year} {1992})}\BibitemShut {NoStop}%
\bibitem [{\citenamefont {Yan}\ \emph {et~al.}(2023)\citenamefont {Yan}, \citenamefont {L\"{u}}, \citenamefont {Chen},\ and\ \citenamefont {Zheng}}]{Yan_2023}%
  \BibitemOpen
  \bibfield  {author} {\bibinfo {author} {\bibfnamefont {Y.}~\bibnamefont {Yan}}, \bibinfo {author} {\bibfnamefont {Z.}~\bibnamefont {L\"{u}}}, \bibinfo {author} {\bibfnamefont {L.}~\bibnamefont {Chen}},\ and\ \bibinfo {author} {\bibfnamefont {H.}~\bibnamefont {Zheng}},\ }\bibfield  {title} {\bibinfo {title} {Multiphoton resonance band and bloch-siegert shift in a bichromatically driven qubit},\ }\href {https://doi.org/https://doi.org/10.1002/qute.202200191} {\bibfield  {journal} {\bibinfo  {journal} {Adv. Quantum Technol.}\ }\textbf {\bibinfo {volume} {6}},\ \bibinfo {pages} {2200191} (\bibinfo {year} {2023})}\BibitemShut {NoStop}%
\bibitem [{\citenamefont {Dicke}(1954)}]{Dicke_1954}%
  \BibitemOpen
  \bibfield  {author} {\bibinfo {author} {\bibfnamefont {R.~H.}\ \bibnamefont {Dicke}},\ }\bibfield  {title} {\bibinfo {title} {Coherence in spontaneous radiation processes},\ }\href {https://doi.org/10.1103/PhysRev.93.99} {\bibfield  {journal} {\bibinfo  {journal} {Phys. Rev.}\ }\textbf {\bibinfo {volume} {93}},\ \bibinfo {pages} {99} (\bibinfo {year} {1954})}\BibitemShut {NoStop}%
\bibitem [{\citenamefont {Zueco}\ and\ \citenamefont {Garc\'{\i}a-Ripoll}(2019)}]{Zueco_2019}%
  \BibitemOpen
  \bibfield  {author} {\bibinfo {author} {\bibfnamefont {D.}~\bibnamefont {Zueco}}\ and\ \bibinfo {author} {\bibfnamefont {J.}~\bibnamefont {Garc\'{\i}a-Ripoll}},\ }\bibfield  {title} {\bibinfo {title} {Ultrastrongly dissipative quantum rabi model},\ }\href {https://doi.org/10.1103/PhysRevA.99.013807} {\bibfield  {journal} {\bibinfo  {journal} {Phys. Rev. A}\ }\textbf {\bibinfo {volume} {99}},\ \bibinfo {pages} {013807} (\bibinfo {year} {2019})}\BibitemShut {NoStop}%
\end{thebibliography}%

\end{document}